# Dynamics of Water Confined in Mesopores with Variable Surface Interaction

Aîcha Jani[a], Mark Busch[b], J. Benedikt Mietner[c], Jacques Ollivier[d], Markus Appel[d], Bernhard Frick[d], Jean-Marc Zanotti[e], Aziz Ghoufi[a], Patrick Huber[b,f,g*], Michael Fröba[c*], and Denis Morineau[a*]

[a]Institute of Physics of Rennes, CNRS-University of Rennes 1, UMR 6251, F-35042 Rennes, France

[b]Hamburg University of Technology, Center for Integrated Multiscale Materials Systems CIMMS, 21073 Hamburg, Germany

[c]Institute of Inorganic and Applied Chemistry, University of Hamburg, 20146 Hamburg, Germany

[d]Institut Laue-Langevin, 71 avenue des Martyrs, F-38000 Grenoble, France

[e]Laboratoire Léon Brillouin, CEA, CNRS, Université Paris-Saclay, F-91191 Gif-sur-Yvette, France

[f]Deutsches Elektronen-Synchrotron DESY, Centre for X-Ray and Nano Science CXNS, 22603 Hamburg, Germany

[g]Hamburg University, Centre for Hybrid Nanostructures CHyN, 22607 Hamburg, Germany

* denis.morineau@univ-rennes1.fr, froeba@chemie.uni-hamburg.de, patrick.huber@tuhh.de

ABSTRACT We have investigated the dynamics of liquid water confined in mesostructured porous silica (MCM-41) and periodic mesoporous organosilicas (PMOs) by incoherent quasielastic neutron scattering experiments. The effect of tuning the water/surface interaction from hydrophilic to more hydrophobic on the water mobility, while keeping the pore size in the range 3.5-4.1 nm, was assessed from the comparative study of three PMOs comprising different organic bridging units and the purely siliceous MCM-41 case. An extended dynamical range was achieved by combining time-of-flight (IN5B) and backscattering (IN16B) quasielastic neutron spectrometers providing complementary energy resolutions. Liquid water was studied at regularly spaced temperatures ranging from 300 K to 243 K. In all systems, the molecular dynamics could be described consistently by the combination of two independent motions resulting from fast local motion around the average molecule position and the confined translational jump diffusion of its center of mass. All the molecules performed local relaxations, whereas the translational motion of a fraction of molecules was frozen on the experimental timescale. This study provides a comprehensive microscopic view on the dynamics of liquid water confined in mesopores, with distinct surface chemistries, in terms of non-mobile/mobile fraction, self-diffusion coefficient, residence time, confining radius, local relaxation time, and their temperature dependence. Importantly, it demonstrates that the strength of the water/surface interaction determines the long-time tail of the dynamics, which we attributed to the translational diffusion of interfacial molecules, while the water dynamics in the pore center is barely affected by the interface hydrophilicity.

## INTRODUCTION

Liquid water is ubiquitous in nature and plays a central role in many applications related to energy and environment. It is noteworthy that in many situations, water is not present as a bulk liquid phase, but rather as a thin interfacial film or as a fluid embedded in a micro-/mesoporous medium. Interestingly, the specific properties of this nanoconfined situation determine the role played by water in many different fields, encompassing the biological activity of proteins,[1, 2] transport through membranes (including biological cells or membranes for proton exchange fuel cells or nanofluidics, nanofiltration/desalinization technology)[3-6] and environments of geological relevance.[7-9]

Mesoporous silica has been extensively used as a model system to study the dynamics of liquid water in spatially confined geometry.[10] Along with other methods such as nuclear magnetic resonance (NMR),[11-13] dielectric spectroscopy,[12, 14] or molecular simulation,[15-17] quasielastic neutron scattering (QENS) has been widely used because it provides a unique viewpoint on the molecular dynamics at the exact space and time scales, which are relevant to nanoconfined liquids (i.e. a few 0.1-1 nm, 0.1-10 ns).[18-26] In general, these studies have revealed a reduction of the water diffusivity compared to the bulk counterpart, which becomes more prominent the smaller the pore size is.

The pioneering studies performed in the mid-90's used Vycor glass.[18] Although the distinction between the surface dynamics and that of the in-pore liquid has been addressed by tuning the filling fraction, additional difficulty stemmed from the interconnected character of the Vycor porosity. The advent of mesostructured porous silica phases, such as MCM-41 and SBA-15, has allowed studying liquid water in well-defined geometry in terms of straight and parallel nanochannels with tunable diameter.[20, 21] These studies made a distinction between two populations of molecules: the water molecules located in the vicinity of the pore walls are immobile or much slower than those located in the inner part of the pores. In addition to this bimodal description, the possible existence of non-exponential relaxations was also discussed and modelled by stretched exponential functions. This approach was intended either to reflect the intrinsic distribution of relaxation times in the liquid dynamics or to reveal the spatially heterogeneous nature of confined fluids.[19, 21, 22]

While MCM-41 and SBA-15 provided adjustable pore size,[20, 26] the evaluation of the effect of surface interaction on the water dynamics was limited due to the unchanged chemical composition of hydrophilic silicas. This issue was addressed by studies, which aimed at disentangling the respective roles of surface interaction and purely geometric effects. Hydrophobic nanopores formed by carbon pores,[27, 28] hydrophobically modified MCM-41,[23] and organosilica phases[25] were used to weaken the surface/water interaction. The other limiting case of ultra-strong interaction was illustrated by the ability of Zr−OH and Al−OH terminated MCM-41 surfaces to immobilize water molecules.[24] A recent study focused on the surface dynamics of water measured at very low coverage fraction of SBA-15 revealed that the non-diffusing fraction of water molecules was involved in slow jump processes between sites that are spatially separated by up to 0.4 nm.[29]

From the above literature, it seems that the chemical nature of the surface of nanoporous materials can be envisioned as a parameter affecting the water dynamics. However, clear-cut conclusions from a comparison of the existing studies are limited by the diversity of instrumental conditions and fitted models. To better address this issue, the present study considers the interfacial mobility of nanoconfined water in the presence of a spatial modulation of the surface/water interaction. This confining condition is realized by using a series of periodic mesoporous silicas (MCM-41)[30, 31] and organosilicas (PMOs).[32-34] These ideal host matrices are formed by a regular triangular lattice of parallel and cylindrical channels, conforming a honeycomb structure. The pore walls of MCM-41 are uniformly formed by silica units with surface H-bond forming silanol groups, serving as a reference hydrophilic matrix. PMOs consist of hydrophobic and (weakly or highly) hydrophilic units, which are alternating along the main pore axis. This is achieved by the introduction of bridging organic units within the silica framework. The molecular-scale periodicity of the arrangement of organic/silica units along the pore axis is demonstrated by (00l) Bragg reflections. Together with the hydrophilic reference pure silica MCM-41, three PMOs comprising different organic bridging units were studied, as illustrated in Table 1 and in Fig. S1. They contain biphenyl (BP-PMO), divinyl-benzene (DVB-PMO), and divinyl-aniline (DVA-PMO) organic bridges, respectively. The four different matrices present comparable mean pore diameters with values between 3.5 and 4.1 nm as detailed in the materials and methods part (Table 1). However, one expects variations in their surface polarity and tendency to form H-bonds, as indicated by water physisorption[35] and multidimensional solid-state NMR spectroscopy studies.[33] Two properties that control the surface hydrophilicity of PMOs have been identified in previous studies. First is the length of the organic bridges that act as spacers between H-bonding silica units. For two PMOs formed by hydrophobic bridging units of different lengths (i.e. benzene and divinylbenzene), it was shown that the surface of the former porous material (B-PMO) was more hydrophilic than the later (DVB-PMO).[35] This was inferred by the lower amount of adsorbed water below capillary condensation and the shift of water pore condensation to higher relative pressure for DVB-PMO. These two phenomena were then used by Mietner *et al.* as parameters to sort the surface hydrophilicity of series of PMOs matrices, carefully selected based on the similarity of their $N_2$ isotherms so that the differences in water adsorption isotherms could be related to surface interaction and not influenced by differences in the size of their pores.[36] A second property that controls the hydrophilicity of PMOs is the capability of the bridging unit to form H-bonds. This phenomenon was observed for DVA-PMO, which presents larger hydrophilicity than BP-PMO due to the tendency of its amino group to form H-bonds with water.[33] 2D $^{1}H$-$^{29}Si$ and $^{1}H$-$^{13}C$ HETCOR NMR experiments showed strong interaction between the adsorbed water and the silicate part of the pore surface for both PMOs.[33] Moreover, additional NMR cross peaks revealed that water also closely interacts with the organic unit for DVA-PMO, which is in contrast to BP-PMO that cannot form H-bonds. Therefore, these systems are well-suited to study the effect of tuning the water/surface interaction from hydrophilic to more hydrophobic by comparing results obtained for DVA-PMO and MCM-41 with respect to DVB-PMO and BP-PMO.

It was recently demonstrated in a pulsed field gradient (PFG) NMR study that the water transport in PMOs actually depends on their chemistry.[34] This method probes the molecule displacements in the μm-range. As a result, the diffusivities were found to be primarily determined by the macroscopic textural properties of the mesoporous particles. Although significant for technological applications, the determination of long-distance transport leaves unanswered questions about the dynamics of water within the pores.

Quasielastic neutron scattering (QENS) is a unique method to resolve the spatio-temporal correlations of water molecule on a timescale ranging from a few picoseconds to a few nanoseconds, and at the nanometer length scale; that is to say a few times smaller than the characteristic distances defined by the pore diameter (in the range 3.5 to 4.1 nm) and by the period of modulation of the surface interaction (about 1.2 nm). We performed incoherent QENS experiments combining two spectrometers: the high flux time-of-flight spectrometer IN5B (ILL, Grenoble) and the high-resolution backscattering spectrometer IN16B (ILL, Grenoble), with respective energy resolution (FWHM) 22 μeV and 0.75 μeV, thus providing complementary observable timescales of about 1 ns and 30 ps. In the explored temperature range (243-300 K), our study demonstrates that the dynamics of confined liquid water can be interpreted by the coexistence of two populations, comprising frozen and mobile molecules on the instrument timescale. We modelled the dynamics of the mobile fraction with two independent relaxations, corresponding to local fast motions and translational jump diffusion. As such, this model conforms to the latest description adopted for bulk water,[37] that was yet to be applied for confined systems. Thanks to the fine control of the surface chemistry of the MCM-41 silica and the three different PMOs, the present work provides a rigorous assessment of the relative role of spatial confinement and surface interaction on the dynamics of confined water. On the one hand, the translational diffusion of interfacial molecules, which is detected in the QENS intensity in the region of small energy transfer, demonstrates a dependence on the surface hydrophilicity. On the other hand, despite clear evidence of confinement effects, this study also concludes on the limited impact of surface chemistry on the dynamics of water molecules located in the center of pores with sizes of 3.5 nm and larger. This conclusion can be reached after a fine and systematic comparison of the dynamic parameters of comparable systems studied under the same conditions.

# METHODS

## Samples

PMO powders were prepared according to the following procedure. NaOH and the alkyltrimethylammonium bromide surfactant were dissolved in deionized water. The bis-silylated precursors of the form $(EtO)_3Si-R-Si(OEt)_3$ (R = organic bridging unit) were added at room temperature and the mixtures were stirred for 20 hours. The mixtures were transferred into a Teflon-lined steel autoclave and statically heated to 95 °C or 100 °C for 24 h. The resultant precipitate was collected by filtration and washed with 200 mL deionized water. After drying at 60 °C, the powder was extracted with a mixture of ethanol and hydrochloric acid (EtOH:HCl (37 %), 97:3, v/v) using a Soxhlet extractor. The porosity and the pore structure of the dried materials were characterized by powder X-ray diffraction (PXRD) and nitrogen physisorption.[34] In this work, we used three different PMO materials with different bridging groups: biphenyl BP-PMO, divinylbenzene DVB-PMO and divinylaniline DVA-PMO. The three bridging units have comparable sizes, which implies that the repetition distances of silica/organic units along the pore axis, as measured by PXRD, are all identical and equal to 1.2 nm, within 1% (cf. Table 1).

The mesoporous materials MCM-41 silicates were prepared in our laboratory according to a procedure similar to that described elsewhere[31] and already used in previous works.[38-42] Hexadecyl ammonium bromide was used as template to get a mesostructured triangular array of aligned channels with mean pore diameter $D$ = 3.65 nm, as confirmed by nitrogen physisorption, transmission electron microscopy (TEM) and neutron diffraction.

The structural parameters of all the matrices are summarized in Table 1. They are comparable, although not exactly the same as those of samples used in previous studies.[33, 34]

**Table 1.** Structural parameters of the mesoporous matrices. They are sorted from top to bottom in the order of decreasing surface hydrophilicity, according to the water sorption analysis by Mietner, performed for a series of matrices with the same surface chemistry but a unique mean pore diameter 3.2 nm.[36]

| Name | Bridging unit name | Bridging unit | Repetition distance of the bridging unit along the pore axis (nm) [a] | Pore volume (cm$^3$.g$^{-1}$) [b] | Specific Surface Area (m$^2$.g$^{-1}$) [b] | Mean pore diameter (nm) [b] | Hydrophilicity [c] |
|---|---|---|---|---|---|---|---|
| MCM-41 | **-** | **-** | **-** | **0.893** | **1077** | **3.65** | **High** |
| DVA-PMO | **Divinyl-aniline** | 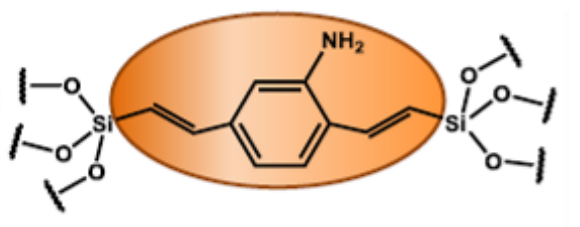 | **1.18** | **0.890** | **1223** | **3.5** | **High** |
| BP-PMO | **Biphenyl** | 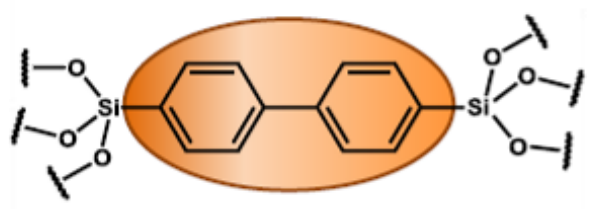 | **1.196** | **0.445** | **786** | **3.5** | **Low** |
| DVB-PMO | **Divinyl-benzene** | 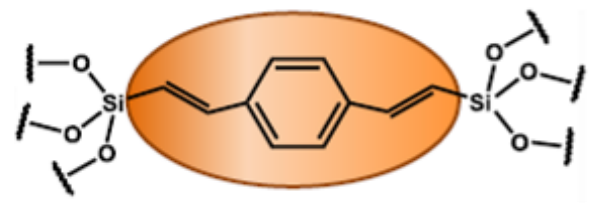 | **1.18** | **0.990** | **864** | **4.1** | **Low** |

[a]: evaluated from the (00l) Bragg reflections. [b]: evaluated from the nitrogen physisorption isotherms. [c]: according to the water sorption analysis by Mietner *et al.*, performed for a series of similar matrices with the same surface chemistry but a unique mean pore diameter 3.2 nm.[36]

All matrices were dried at 120 °C under a primary vacuum. To prepare hydrated matrices, a constant amount of mesoporous materials was charged in a flat alumina rectangular cell (1 mm of thickness) and placed in a desiccator in the presence of a beaker containing a saturated aqueous solution of NaCl. The resulting relative humidity (RH) of 75% at 25 °C was above the partial pressure of the capillary condensation[33] and allowed the complete filling of the pore system. The sample was kept in the atmosphere of constant humidity for 24 h to ensure that an equilibrium condition is reached. The cells were sealed with an indium wire to avoid water loss and ensure a constant hydration level during neutron scattering experiments. We also conducted differential scanning calorimetry (DSC) experiments on capillary filled samples that were saturated with water at the same RH, as detailed in the Supporting Information. These measurements have confirmed that this filling procedure led to the capillary filling of the pores with no detectable bulk excess water outside the porous matrix. They provided a more accurate determination of the resulting water mass fraction. A second set of samples filled at a RH of 100% was used to determine an upper limit of the water mass fraction at which overfilling occurs. Finally, we could also conclude on the absence of crystallization in the temperature range used for QENS experiments.

**QENS experiments**

Quasielastic neutron scattering experiments were carried out using two spectrometers with different energy resolutions at the Institut Laue Langevin (ILL, Grenoble, France).[43, 44] The disc chopper time-of-flight spectrometer IN5B was used with an incident wavelength of 8 Å. In this configuration, the resulting energy resolution $\Delta E$ around the elastic peak is about 22 µeV (FWHM), corresponding to a timescale ($t=\Delta E/\hbar$) of about 30 ps. The quasielastic signal retained for the data evaluation covered an energy range ($E=\hbar\omega$) between -5.0 and 0.7 meV and a $Q$ range between 0.2 and 1.3 Å$^{-1}$. The high-resolution IN16B spectrometer was chosen with unpolished Si (111) monochromator and analyzers in backscattering geometry, which corresponds to an incident wavelength of 6.271 Å and results in an energy resolution of 0.75 µeV and a corresponding timescale of about 1 ns. The energy range was ±30 μeV with a $Q$ range between 0.19 and 1.83 Å$^{-1}$. The background chopper of the instrument was run in its high signal-to-noise mode.[45] A cryofurnace and an ILL orange cryostat, were used on IN16B and IN5B spectrometers in order to regulate the sample temperature. The measurements were performed after thermal equilibration at regularly spaced temperatures, which were reached sequentially on cooling (300, 278, 258 and 243 K on IN5B, and 278, 258 and 243 K on IN16B) with dried and water filled samples. The four different matrices were studied on IN5B, and three of them on IN16B (MCM-41, DVB-PMO, and DVA-PMO).

Standard data corrections were applied using the packages MANTID[46] and LAMP[47] provided at the ILL. The experimental intensity was corrected for detector efficiency, for the background contribution arising from the empty cell and spectrometer, and transformed into the $Q$ and energy dependent scattering function $S(Q,\omega)$. The fitting of scattering functions $S(Q,\omega)$ in the frequency domain was carried out using the QENSH program provided by the Laboratoire Léon Brillouin (LLB, Saclay, France).

## RESULTS and DISCUSSSION

### Data modelling

The QENS spectra of water filled DVB-PMO and MCM-41 are shown in Fig. 1 and Fig. 2. Qualitatively similar results were obtained for water confined in the other matrices (Fig. S4 and Fig. S5). The temperature dependence of the scattering intensity is illustrated in Fig. 1 for the selected value of the transfer of momentum $Q$ = 0.8-0.85 Å$^{-1}$. We observed a quasielastic broadening that covered a typical energy range from about 0.1 to 1 meV on IN5B and 30 μeV on IN16B. Given the dominant contribution from the incoherent scattering cross section of hydrogen atoms, this quasielastic signal is attributed to the self-part of the dynamic structure factor $S(Q,\omega)$. Thus, it provides information on the single particle dynamics of water molecules. A continuous sharpening of the quasielastic peak was observed on decreasing the temperature from 300 K to 243 K. On a qualitative level, the width of the quasielastic scattering is inversely proportional to the typical timescale of motion of particles. Therefore, this sharpening indicates that the water dynamics slows down gradually on cooling. The absence of a sharp increase of the elastic signal confirms that water remained liquid on the entire studied temperature range. It allows excluding the hypothetical presence of water outside the pores and the crystallization of in-pore water, which is in agreement with the DSC results. The $Q$-dependence of the scattering intensity is illustrated in Fig. 2 for IN5B at the temperature $T$ = 278 K. On increasing $Q$, we observed a significant reduction of the intensity contained in the elastic resolution (about a factor of ten) combined with a broadening of the quasielastic line. This $Q$-dependence probably arises from the combination of different effects, which are classically observed in QENS studies of liquids. It includes the effect of dispersive modes (e.g. translational diffusion) and the modulation of the elastic intensity due to vibrations (Debye Waller factor) or local quasielastic relaxations (e.g. rotation, libration), which comprise an elastic incoherent structure factor (EISF).

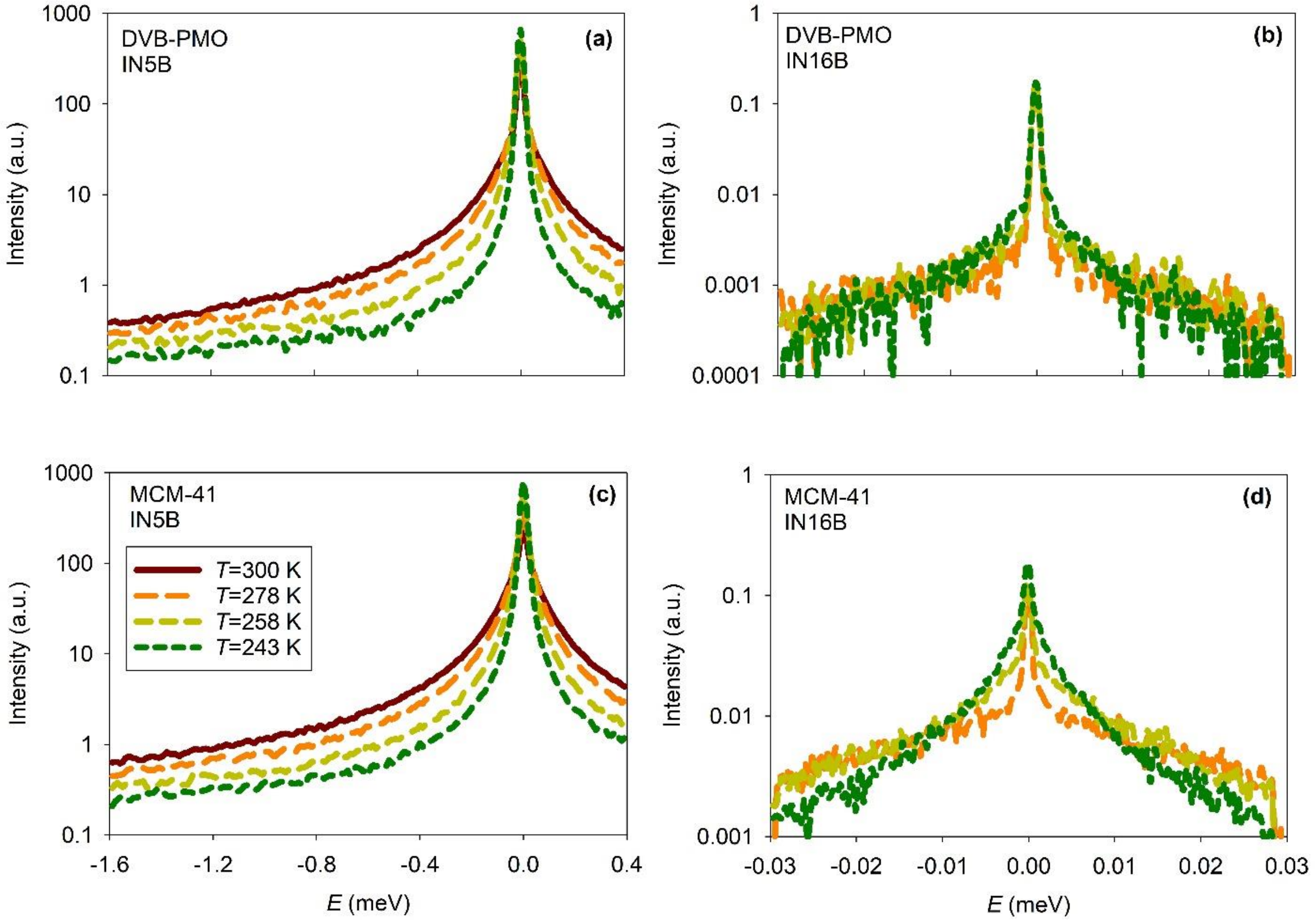

**FIG. 1.** Temperature dependence of the scattering intensity of water filled DVB-PMO (upper panels a, b) and water filled MCM-41 (lower panels c, d). QENS spectra measured on IN5B at $Q = 0.8$ Å$^{-1}$ (left panels a, c) and on IN16B at $Q = 0.85$ Å$^{-1}$ (right panels b, d).

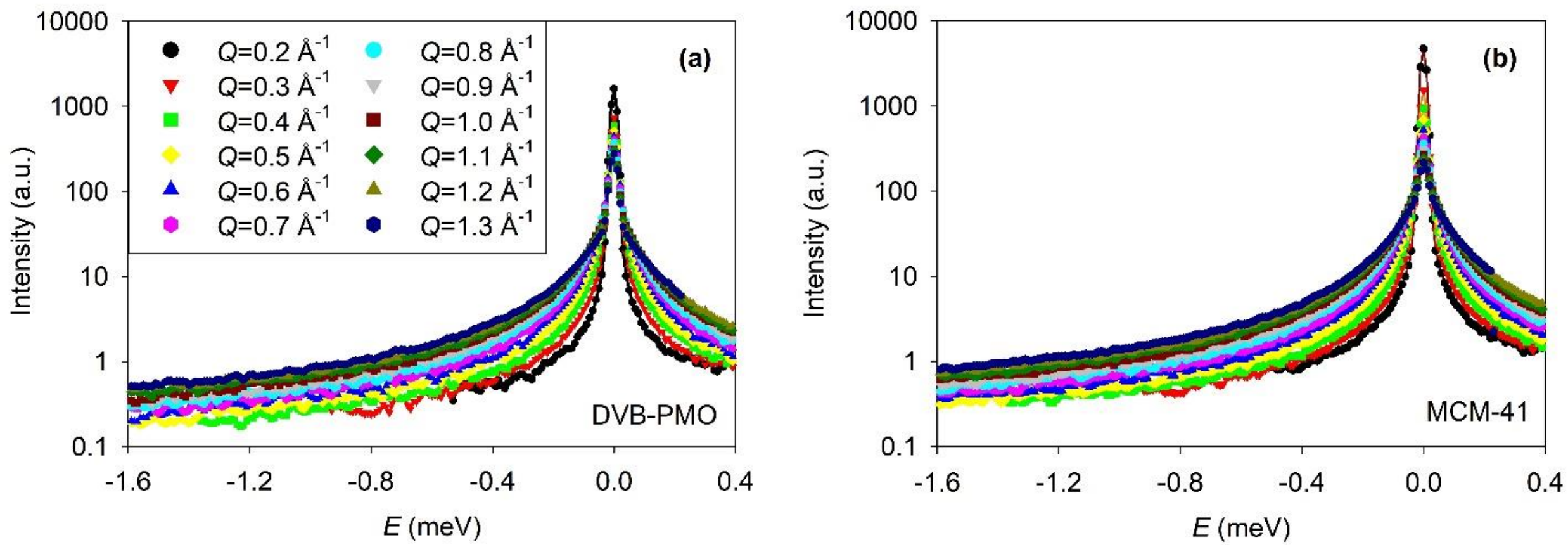

**FIG. 2.** $Q$-dependence of the scattering intensity measured on IN5B at the temperature $T$ = 278 K of water filled DVB-PMO (left panel a) and water filled MCM-41 (right panel b).

All spectra were fitted individually at each $Q$ with a model comprising one elastic component, i.e. a Dirac function $\delta(\omega)$, and a quasielastic contribution, which was approximated by one (IN16B) or a sum of two (IN5B) Lorentzian functions, as described in SI. The essentially elastic contribution arising from the pore wall atoms was accounted for by adding the scattered experimental intensity of the empty matrices to the theoretical functions, which were fitted to the experimental spectra of the water filled matrices. This means that the fitted Dirac function can be actually ascribed to an elastic contribution of water molecules. Following the assignment made in recent studies of bulk water, we related the broad and sharp quasielastic components to two distinct dynamics: local motion (L) and translational diffusion (T), respectively.[37, 48] It is likely that only the slowest one, attributed to translation, was actually detected on IN16B. In fact, the broad quasielastic contribution arising from local dynamics likely appeared as a vanishing flat background, due to the reduced energy range covered by the high-resolution instrument. Therefore, the incoherent dynamic structure factor of water writes as:

$$S(Q,\omega) = K\,[A_T(Q)\delta(\omega) + (1 - A_T(Q))L_T(Q,\omega,\Gamma_T)]\otimes[A_L(Q)\delta(\omega) + (1 - A_L(Q))\,L_L(Q,\omega,\Gamma_L)] \quad (3)$$

where $K$ is a scaling factor comprising the attenuation of the scattering intensity due to inelastic vibrational modes (Debye-Waller factor). The functions $A_T(Q)$ and $A_L(Q)$ are the elastic incoherent structure factors (EISFs) of the translational and local motions.[49] Their respective linewidth $\Gamma_T$ and $\Gamma_L$ were effectively separable since they differ by a factor of 7-10. These quantities were evaluated from the fitted parameters introduced in Eqs. (1) and (2) after straightforward calculations, and considering that $\Gamma_T \ll \Gamma_L$.

**Local water dynamics**

The experimental EISFs $A_L(Q)$ obtained from the fitting of Eq. (3) are illustrated in Fig. 3. They present an important elastic contribution, which is above 60% at 300 K and 95% at 243 K on the studied $Q$-range. It demonstrates the non-diffusive (i.e. localized) character of this motion. In the classical approach of Teixeira *et al.*[50] this dynamics was described as an isotropic rotation of hydrogens around the oxygen atom. The corresponding EISF writes as $A_R(Q) = [j_0(QR_R)]^2$, where $j_0$ is the zeroth-order spherical Bessel function and $R_R$ = 0.98 Å for the O-H bong length.[49] A comparison of the prediction from the classical approach with the EISFs $A_L(Q)$ determined experimentally for water in MCM-41 is illustrated Fig. 3(a). Only at 300 K, the standard model agreed well with the experiments. However, this model was unable to describe the increase of intensity observed at lower temperature. Satisfactory fits could still be obtained with the same functional form $A_R(Q)$, provided that $R_R$ was allowed to vary with the temperature. When considered as a free fit parameter, the value of $R_R$ decreased with decreasing temperature. It reached values as small as 0.5 Å, which cannot be related to the O-H bond length anymore. In that sense, the model lost its physical meaning. Another approach that maintains the classical interpretation of $A_L(Q)$ consists in adding a supplement $Q$-independent elastic term to the isotropic rotation function $A_R(Q)$ (i.e. $A_L(Q) = p\delta(\omega) + (1-p)A_R(Q)$). From this viewpoint, it would imply that a fraction $p$ of the water molecules does not perform rotation on the instrumental timescale due to confinement and interfacial restrictions. At 300 K, such an additional elastic peak was not required to best fit the data, meaning that all the molecules were dynamically active. However, this alternative model led to a large elastic component at sub-ambient temperature, reaching up to 80% at the lowest temperature. It is unlikely that the rotational diffusion would be frozen for such a large fraction of molecules, while it would be active and very fast (i.e. about 1-3 ps) for the unfrozen ones. It is also inconsistent with the analysis of the translational diffusion detailed in the next section, where much smaller fractions of dynamically frozen molecules were obtained. Our results rather suggest that this broad quasielastic line does not strictly correspond to isotropic rotational diffusion. This conclusion appears in line with an alternative view proposed by Qvist *et al.* for bulk water.[37] This motion was attributed to a local dynamical “intra-basin” relaxation, also observed in the water molecules trajectories of Molecular Dynamics (MD) simulations. A comparison is made in Fig. 3(a) with the polynomial fit of the bulk water data.[37] It is similar to the EISFs that we measured at 258 K. A univocal determination of the geometry of this local relaxation is elusive, as also discussed for bulk water by Arbe *et al*.[48] However, based on a Debye-Waller factor type of analysis, we can conclude on the gradual shrinkage of the mean-squared-amplitude of the local motion on decreasing the temperature for confined water.

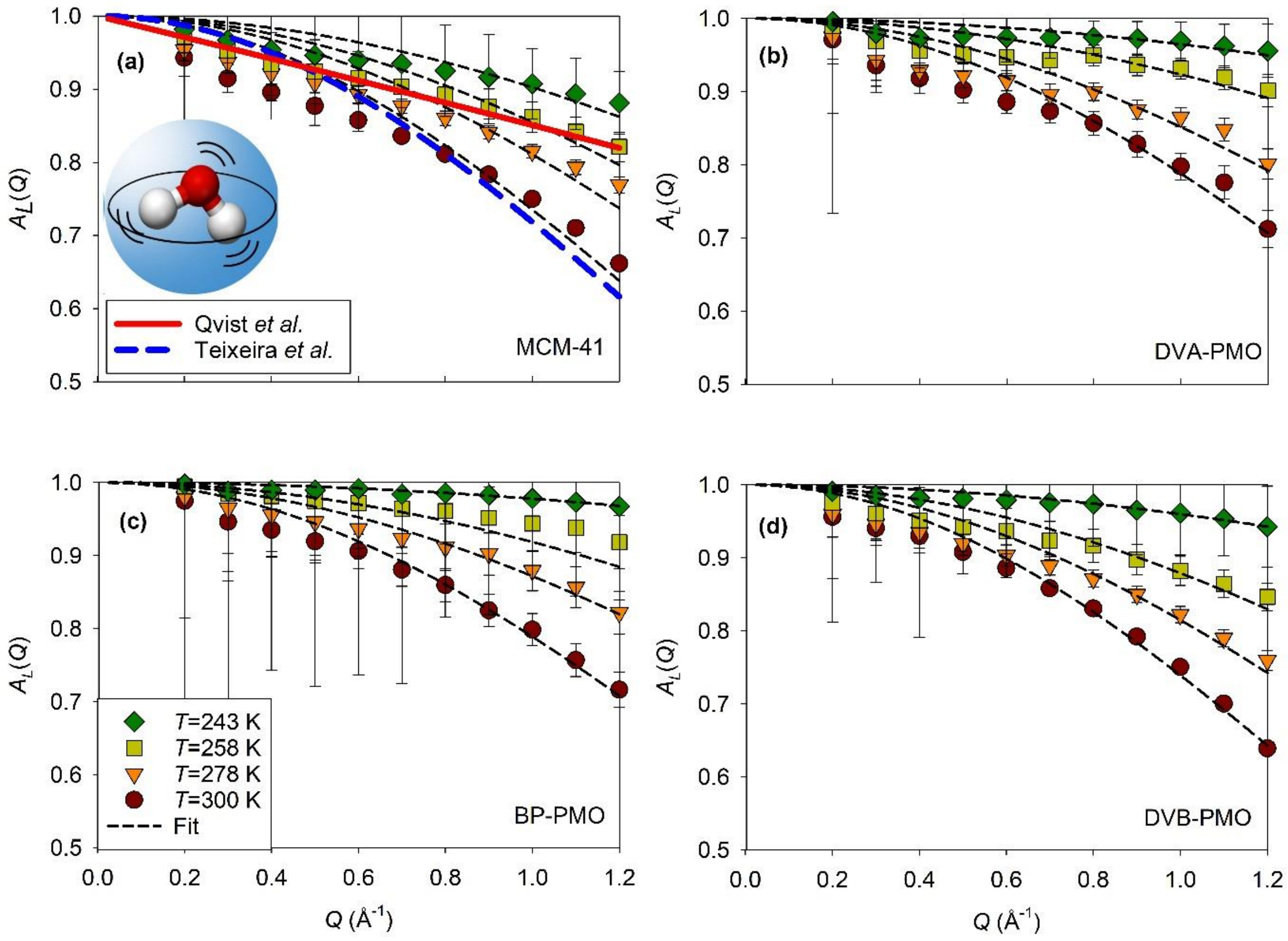

**FIG. 3.** Elastic incoherent structure factor of the local quasielastic relaxation (symbols) deduced from IN5B measurements at four different temperatures of water confined in different matrices (a) MCM-41, (b) DVA-PMO, (c) BP-PMO, and (d) DVB-PMO. Two different approaches describing the broadest quasielastic relaxation of bulk water are presented for comparison in the panel (a): (solid line) model-free EISF of the local relaxation averaged over 6 temperatures in the range 253-293 K, after Qvist *et al*.[37] and (long dashed line) the EISF of the continuous molecule rotation supposed independent on the temperature for the same temperature range after Teixeira *et al.*[50] Inset of panel (a): sketch of the local motion of water.

On the entire temperature range studied, the relaxation time $\tau_L$ attributed to the local motion of water varied from 1 to 3 ps. These values are within the range of relaxation times reported for bulk water (1 to 3.5 ps) on a slightly reduced temperature range.[37] The effect of the matrix can be estimated from a comparison of the EISFs measured for water confined in MCM-41 (cf. Fig. 3(a)) and the three PMOs (cf. Figs. 3(b-d)). If present, the difference between the EISFs is very small and hardly resolved within experimental uncertainties. The temperature dependence of the relaxation time of water confined in the four different matrices was also very similar, with an averaged value of the activation energy of $10 \pm 4$ kJ·mol$^{-1}$. The previous QENS studies of confined water were often limited to translational diffusion and results about the fast local dynamics are scarcer in the literature. To the best of our knowledge, there exists only one available QENS measurement of water confined in a benzene-bridged PMO by Aso et al.[25] In this study, the authors concluded on a significant slowdown (by a factor of three) of the local dynamics of water confined in MCM-41, with respect to those of water confined in the PMO and bulk water. We note that the comparison with this PMO's result was based on an older QENS measurement of water confined in MCM-41,[20] which was moreover performed at two different temperatures. On the contrary, our conclusions are based on a single study, with measurements performed systematically under the same experimental conditions. Under these conditions, our results rather point towards a comparable fast local dynamics of water when confined in PMOs or in MCM-41. It is also worth mentioning that this conclusion is in line with the pioneering study of water in Vycor, which concluded that the fast QENS relaxation denoted as 'hydrogen-bond lifetime' was very close to the bulk one, both for fully and partly hydrated systems.[18]

## Translational water dynamics – neutron time-of-flight study

The sharpest quasielastic component detected on IN5B was attributed to the translational diffusion of the molecules' center of mass. However, we noticed that this intensity also contained a purely elastic component, as illustrated in Fig. 4. The existence of an EISF associated to this component, of which the intensity was dependent on both the temperature and the momentum transfer $Q$, is inconsistent with long range translation diffusion. The EISFs were fitted with a model that incorporate a fraction $p$ of non-mobile molecules (*i.e.* dynamically frozen on the timescale of the instrument) and a fraction (1-$p$) of molecules performing restricted translational diffusion in a sphere.[49, 51]

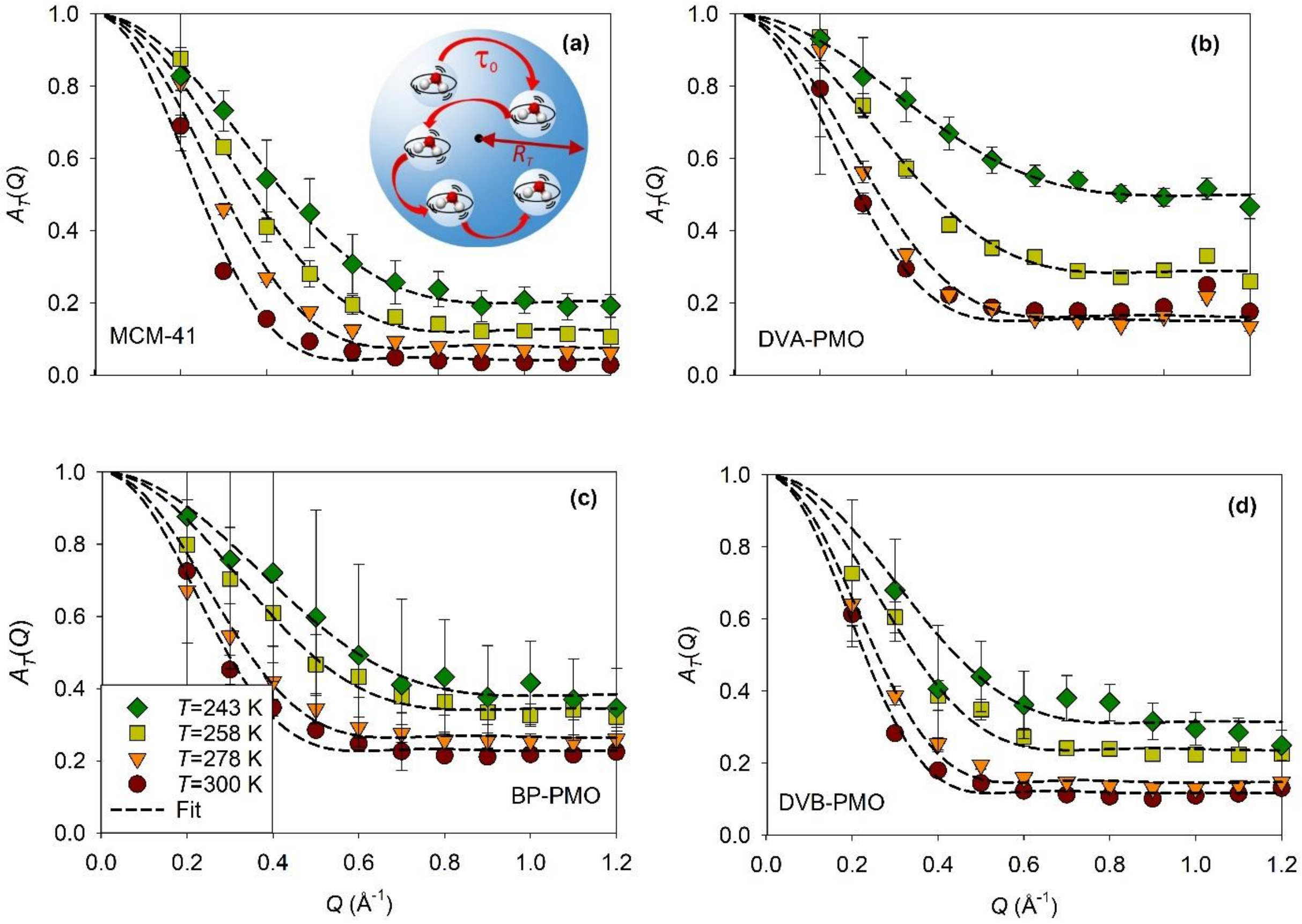

**FIG. 4.** Elastic incoherent structure factor of the first quasielastic relaxation (symbols) deduced from IN5B measurements at four different temperatures of water confined in different matrices (a) MCM-41, (b) DVA-PMO, (c) BP-PMO, and (d) DVB-PMO. Fits with a model of translation diffusion confined in a sphere (lines). Inset of panel (a): sketch of the confined jump diffusion motion of water.

The corresponding expression of $A_T(Q)$ is

$$A_T(Q) = p\delta(\omega) + (1-p)\left[\frac{3j_1(QR_T)}{QR_T}\right]^2 \quad (4)$$

where $R_T$ is the radius of the confining sphere, $p$ the fraction of non-mobile molecules, and $j_1$ is the first-order spherical Bessel function. This model could reproduce the experimental EISFs quantitatively, as illustrated by dashed lines in Fig. 4. However, the parameters obtained at the lowest temperatures ($T$ = 243 K) presented large uncertainties, which was due to difficulties in separating the sharpest quasielastic component from a pure elastic one within the energy resolution of IN5B. The two fitted parameters $p$ and $R_T$ showed a systematic temperature dependence as can be inferred from Fig. 5. A systematic increase of $p$ with decreasing temperature reflects the larger number of dynamically frozen molecules on cooling. It is about 10-20% at 300 K, and reaches values as large as 40-50% at 243 K. We also noticed a dependence of $p$ on the confining matrix (cf. Fig. 5(a)). The smallest value of $p$ was obtained for the purely siliceous matrix MCM-41. It corresponds to half the value obtained for BP-PMO that comprises hydrophobic bridging. At first sight, this feature appears counterintuitive. Indeed, MCM-41 has a larger surface density of silanol groups that may act as adsorbing sites, thus reducing the translational motion of water. On the other hand, the barrier of activation related to the molecule jump from one surface site to an adjacent site can be reduced when the surface presents a homogeneous distribution of equivalent silica units such as for MCM-41. For PMOs instead, the alternation of hydrophobic and hydrophilic sites might interrupt the surface diffusion of water. This reasoning should be moderated for hydrophilic DVA-PMO, considering that the amino groups also act as H-bonding sites, as shown by NMR.[33] The distance between $NH_2$ and silanols that is estimated to be a bit less than half of the repetition distance (i.e. 5 Å) is still too large to allow a direct interaction of a water molecule with both sites. We believe that this issue needs additional support from other methods, such as molecular simulation.

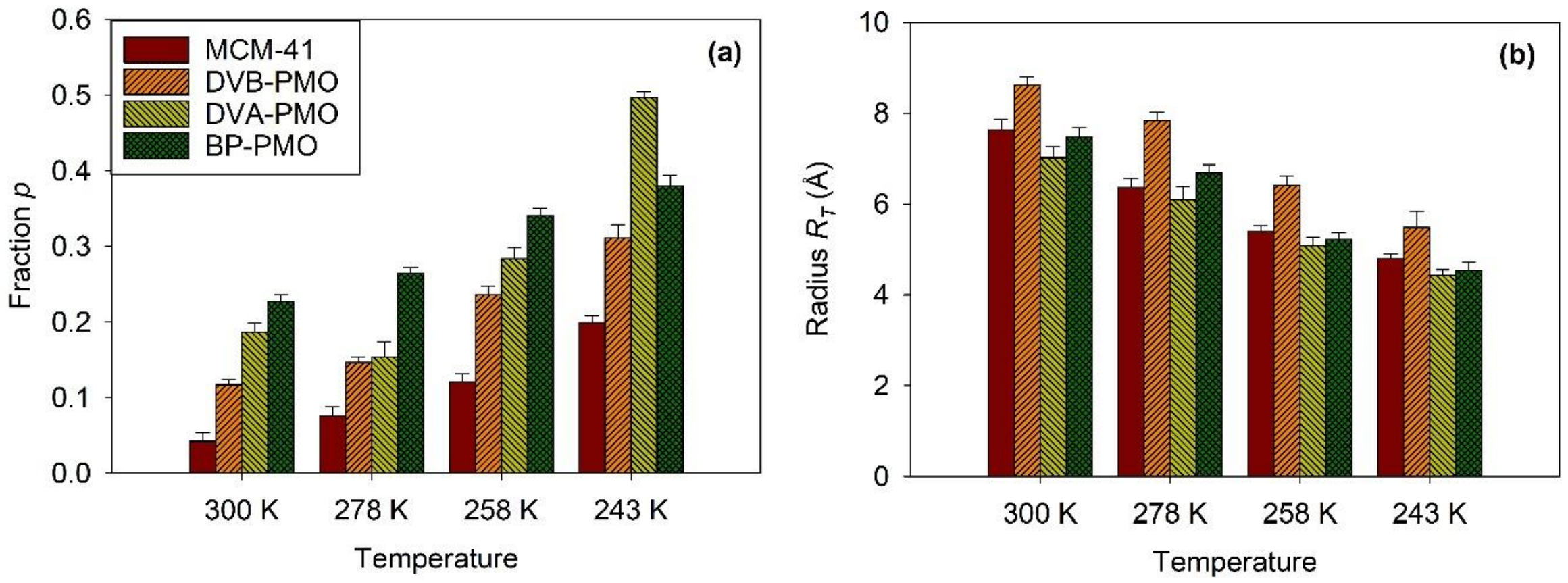

**FIG. 5**. Variation with the temperature and the confining matrix of the parameters obtained from the fit of the EISF of the translational diffusion (from IN5B measurements). (a) Fraction *p* of dynamically frozen water molecules. (b) Radius $R_T$ of the sphere confining the translational diffusion.

The radius $R_T$ of the sphere confining the translational diffusion of the fraction (1-*p*) of mobile molecules decreases from 9 to 5 Å, during cooling from 300 to 243 K (cf. Fig. 5(b)). The spatial restriction of the translational displacements on cooling is similar to the observation made for the local dynamics in the previous part in that it is decreasing. The values of $R_T$ spanned from one quarter to one half of the pore radius. Hence, this confining sphere cannot be strictly identified as the pore itself. However, a correlation between $R_T$ and $R_{pore}$ is suggested by the systematically larger values of $R_T$ for DVB-PMO, which pore size is 15% larger than for the three other matrices (cf. Table 1). This indicates a possible link between $R_T$ and geometrical aspects of confinement. Moreover, the role of surface effects can be invoked. Indeed, the fraction $p$ of non-mobile molecules could be also taken into account as a secondary source of confinement. For simplicity, it can be assumed that these molecules are forming a homogeneous layer located at the interface with the pore wall. In this case, the radius $R_{core}$ of the core region of the pore, where the mobile water molecules are located, is $R_{core} = R_{pore}\sqrt{(1-p)}$. Using the experimental values of $p$ and $R_{pore}$ from Table 1, we obtained that $R_{core}$ varies from 19 to 12.5 Å as a function of the matrix and the temperature, which is about twice the value of $R_T$. Interestingly, the thickness of the non-mobile layer (*i.e.* $e_{non-mob} = R_{pore} - R_{core}$) is in the range 1-5 Å, and is maximum at the lowest studied temperature 243 K, before crystallization occurs. It compares reasonably with the thickness of the non-crystalizing layer ($t$ = 6 Å) evaluated by cryoporometry for water confined in MCM-41 and SBA-15 porous silicates, [52, 53] as well as with the thickness of the immobile, sticky layer inferred from capillary rise experiments for water in hydrophilic silica pores (Vycor).[54]

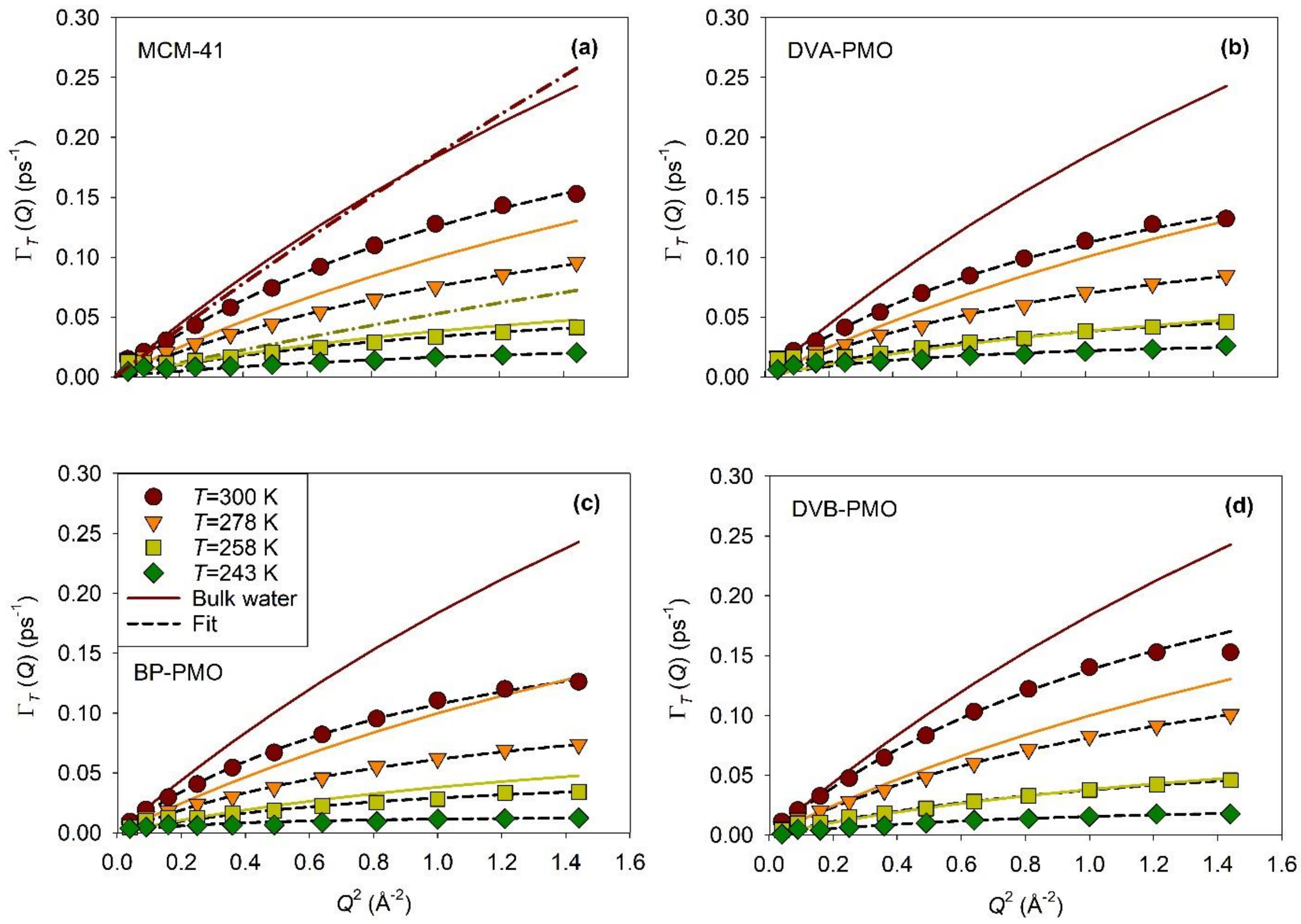

**FIG. 6**. Evolutions of the half width at half-maximum $\Gamma_T$ of the sharp Lorentzian as a function of $Q^2$ obtained from the fitting of QENS spectra measured on IN5B for water confined in the four matrices (a) MCM-41, (b) DVA-PMO, (c) BP-PMO, and (d) DVB-PMO at four temperatures (filled symbols). Fit using the jump diffusion model (thin dashed lines). Bulk water experimental results at three temperatures, from top to bottom 300 K, 278 K, and 258 K (solid lines) from Teixeira *et al.* [50] In panel (a), bulk water experimental results at 293.4 K, and 257.8 K (dashed dotted line) from Qvist *et al*.[37]

After the discussion of the EISFs, we address now the corresponding quasielastic part, which contains dynamical information about the fraction (1-*p*) of mobile molecules. The linewidth of the sharper Lorentzian quasielastic line $\Gamma_T$ found on IN5B is illustrated in Fig. 6, as a function of the squared momentum transfer. The different panels correspond to the four different matrices, and the different curves illustrate different temperatures. The experimental results for bulk water are also illustrated for comparison. Only two curves from Qvist *et al*.[37] are presented in Fig. 6a, because, unlike for the study by Teixeira *et al.* [50], the selection of temperatures was different from the one used the present work. At intermediate $Q$s, $\Gamma_T$ increases linearly with $Q^2$, which conforms the normal Fick's law of translational diffusion. However, deviations occur both at small and large $Q$s. When $Q$ tends to zero, $\Gamma_T$ does not really vanish but rather saturates towards a finite value, as illustrated in Fig. S7 for MCM-41. This feature can be explained as resulting from the confined character of the translational diffusion, which was demonstrated above from the analysis of the EISFs. In the frame of the model of confined diffusion in a sphere, it was predicted by Volino *et al*.[51] that $\Gamma_T$ should deviate from the Fick law at $Q \approx \frac{3.3}{R_T}$ towards a plateau value $\Gamma_T = \frac{4.33\, D_T}{{R_T}^2}$. The measured linewidths are in fair agreement with this prediction, as illustrated in Fig. S7 by the shaded areas. At large $Q$s, $\Gamma_T$ bends and tends asymptotically towards a constant value denoted $1/\tau_0$ (Fig. 6). In fact, the Fickian diffusion model assumes a continuous motion process. Deviation from this assumption is observed when considering small displacements (*i.e.* for $Q$ larger than the inverse particle distance), where a discontinuous mechanism is related to the finite molecular size and the local order in the liquid. The linewidth was modelled by the well-known jump-diffusion model, which assumes that the translation motion proceeds by successive elementary jumps (thin dashed lines in Fig. 6).[49, 55] Between two jumps, the particle remains localized for a typical residence time $\tau_0$ on a molecular site, with a spatial extension limited to the amplitude of vibrational modes. Applying the usual assumptions that the jump can be regarded as instantaneous with respect to the residence time $\tau_0$ spent by the particle on a site, and that the jump length $l$ is much larger than the spatial extension of each site, the linewidth of the Lorentzian was fitted with

$$\Gamma_T(Q) = \frac{D_T Q^2}{1+\tau_0 D_T Q^2} \tag{5}$$

where $D_T$ is the diffusion coefficient and $\tau_0$ the mean residence time.

These two parameters obtained for the four different matrices are illustrated in Fig. 7 as a function of the temperature in Arrhenius coordinates. The bulk values of $D_T$ and $\tau_0$ determined from QENS measurements,[18, 50] and the diffusion coefficient measured by NMR[56] are also added for comparison. The values of the diffusion coefficient for the four confined materials are, within about 10%, in good accordance with those of the bulk water. Deviations observed at high and low temperature could indicate that the activation energy is reduced in confinement, although the difference remained in the limit of the data accuracy, especially at 243 K as $\Gamma_T$ approaches the instrumental resolution limit of IN5B. Significant differences between the confined and bulk water were obtained for the residence time $\tau_0$. It is systematically larger in confinement than in the bulk as illustrated in Fig. 7(b). At 300 K, the residence time is three to four times longer in the porous matrices. Moreover, the activation energy is smaller in confinement, so that the residence times of the different systems apparently approach the bulk system at lower temperature (243 K). It is worth noting that both $D_T$ and $\tau_0$ exhibit a curvature, which indicates a super-Arrhenius character as usually observed in supercooled glass-forming liquids.[57, 58] This is illustrated in Fig. 7(b) by the fit of the residence time with a Vogel-Fulcher-Tammann equation (dashed line). A mean jump length $l$ could be estimated by combining the diffusion coefficient and the residence time noting that $D_T = \frac{\langle l^2 \rangle}{6\tau_0}$. The obtained values of $l$ were in the typical range from 0.65 to 1.2 Å for all the matrices with a systematic reduction when the temperature decreased from 300 to 243 K.

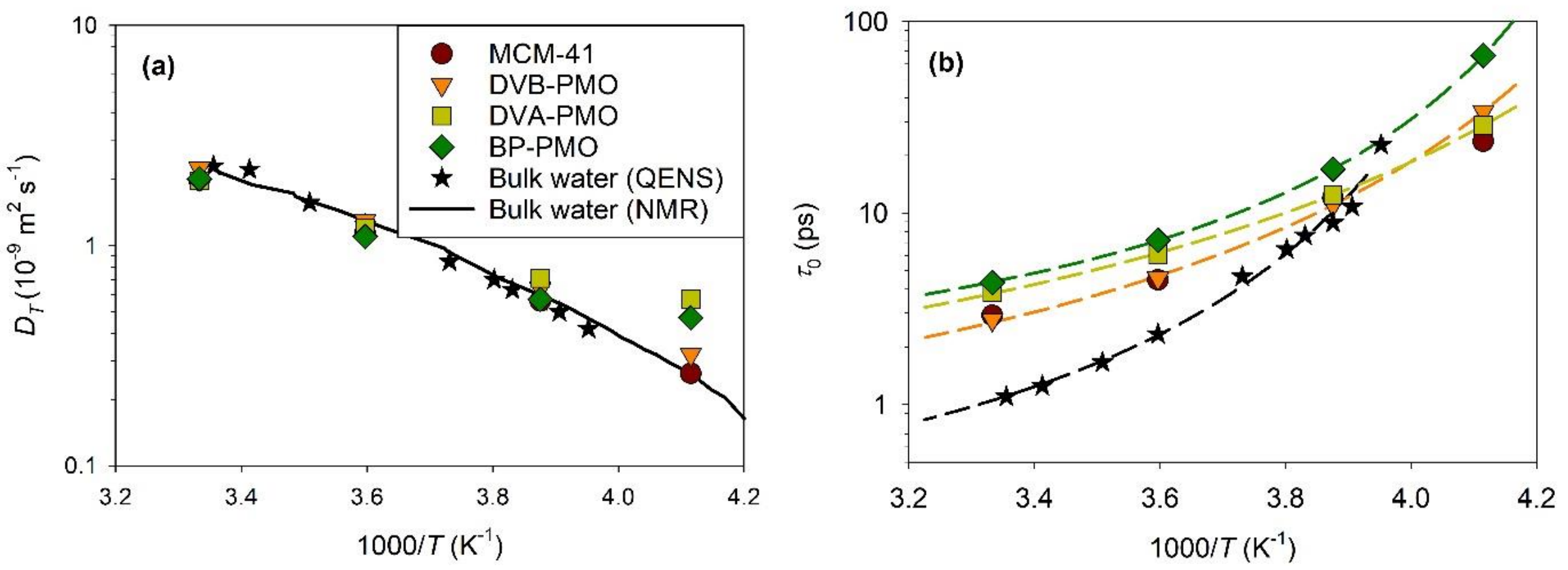

**FIG. 7**. (a) Translational diffusion coefficient $D_T$ and (b) residence time $\tau_0$ of water evaluated from the fit of IN5B spectra with a jump-diffusion model for water confined in MCM-41 and three PMOs as a function of the inverse temperature. In panel a, the QENS data of bulk water are from Teixeira *et al.*[50] (stars) and the NMR diffusion coefficient from Price *et al.*[56] (solid line). VTF fits of the residence time (dashed line in panel b).

A comparison with other QENS data obtained from the literature is given in Table 2. These experiments have been analyzed differently, and often the EISFs related to the fast local dynamics, the confined translation diffusion and the fraction of non-mobile molecules were not available simultaneously. Also, the temperature range was sometimes limited, which prevented a thorough comparison with our results. Bearing in mind these possible limitations, it is, however, possible to compare the values of $D_T$ and $\tau_0$ that were obtained from the quasielastic linewidth. Similar conclusions can be made from these previous evaluations of $D_T$, which values remain similar or marginally smaller than that of the bulk water whenever the pore diameter is larger or equal to 3 nm. An enhanced slowdown of the translational dynamics is observed for a pore size smaller than 3 nm (Table 2).

**Table 2.** Translation diffusion coefficient $D_T$ and the residence time $\tau_0$ of the mobile fraction of water molecules derived from the jump diffusion model. The results from the present study are compared with those from the literature.

| **Temperature (K)** | **$D_t$ ($10^{-9}$ $m^2.s^{-1}$) $\tau_0$ (ps)** | **MCM-41 3.6 nm** | **DVA-PMO 3.5 nm** | **BP-PMO 3.5 nm** | **DVB-PMO 4.1 nm** | **Bulk[a]** | **Ph-PMO[c] 3 nm** | **SBA-15[d] 6.6 nm** | **MCM-41[e] 3.8 nm** | **MCM-41[d] 2.9 nm** | **MCM-41[d] 2.4 nm** |
|---|---|---|---|---|---|---|---|---|---|---|---|
| **300** | $D_t$ | 1.98 | 1.97 | 2.01 | 2.24 | 2.3[b] (298 K) | 1.7 | 2.0 | 1.7 | 1.5 | 1.2 |
| | $\tau_0$ | 2.9 | 3.9 | 4.3 | 2.8 | 1.1[b] (298 K) | 2.6 | 3.3 | 2.6 | 5.3 | 7.3 |
| **278** | $D_t$ | 1.14 | 1.20 | 1.10 | 1.30 | 1.30 | 1.1 (273 K) | | 0.77 (273 K) | | |
| | $\tau_0$ | 4.5 | 6.1 | 7.2 | 4.6 | 2.3 | 8.7 (273 K) | | 6.6 (273 K) | | |
| **258** | $D_t$ | 0.56 | 0.71 | 0.57 | 0.64 | 0.56 | | | | | |
| | $\tau_0$ | 12.0 | 12.4 | 17.0 | 11.0 | 8.9 | | | | | |
| **243** | $D_t$ | 0.26 | 0.57 | 0.47 | 0.32 | 0.42 (253 K) | | | | | |
| | $\tau_0$ | 23.7 | 28.6 | 66.6 | 33.9 | 22.7 (253 K) | | | | | |

a: data from Teixeira *et al.*[50], b: data from Bellissent-Funel *et al.*[18], c: data from Aso *et al.*[25], d: data from Baum *et al.*[26], and e: data from Takahara *et al.*[20]

**Interfacial translational water dynamics - high resolution neutron backscattering study**

In the previous section, the translational dynamics of confined water has been discussed in terms of two populations: an interfacial layer of thickness $e_{non-mob}$ in the range 1-5 Å, with a highly reduced dynamics due to the interaction with the pore surface, and a liquid region at the pore center with a 'bulk-like' dynamics with moderate slowdown and marginal effect of the pore chemistry. It is noteworthy that the interfacial dynamics is likely not frozen on a longer time scale. The fraction of molecules named 'non-mobile' rather corresponds to the long-time tail of the broad distribution of dynamics.[58] This simplified bimodal description is widely used to account for the dynamics that is slower than the cut-off introduced by the energy resolution of the instrument. To get a better insight on the dynamics of the interfacial molecules, which are expected to be more influenced by the nature of the water/surface interaction, we have used the higher resolution of the neutron backscattering spectrometer IN16B, i.e. 0.75 μeV compared to 22 μeV on IN5B. The resulting observable timescale is thus extended from about 30 ps to 1 ns. In the meantime, the dynamical range covered on IN16B is much reduced, which means that the broad quasielastic intensity shown for IN5B is mostly hidden in the background on IN16B (i.e. the local motion and possibly the fastest part of the translation).

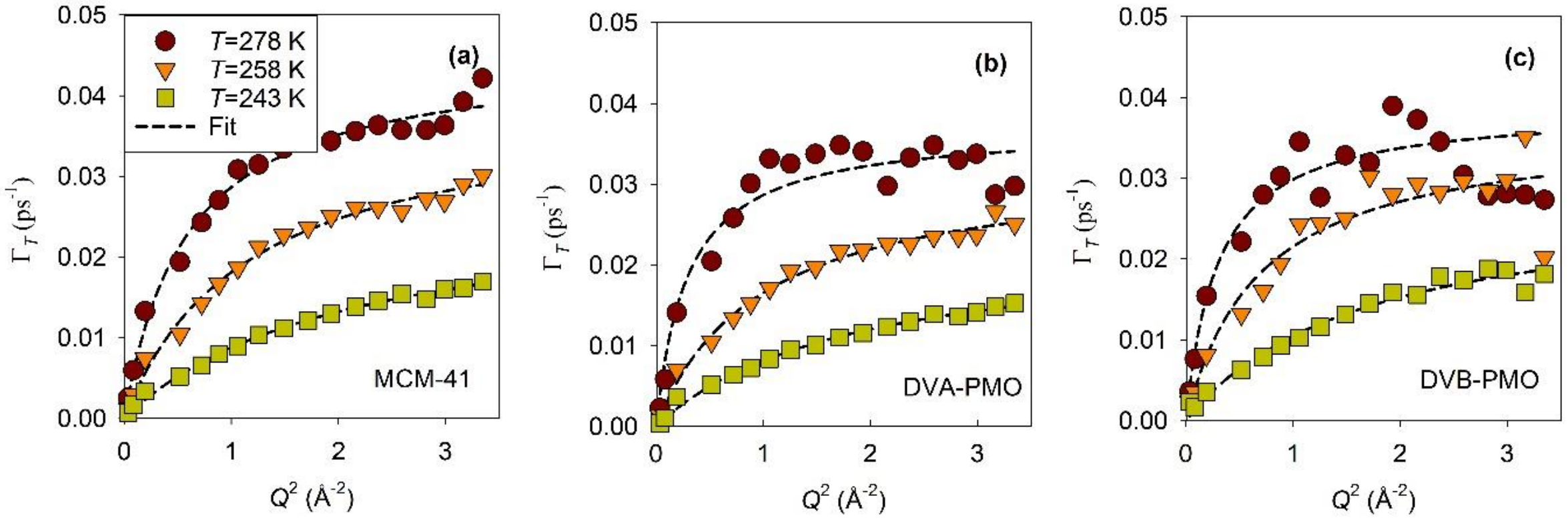

**FIG. 8**. Evolutions of the half width at half-maximum $\Gamma_T$ of the Lorentzian as a function of $Q^2$ obtained from the fitting of QENS spectra measured on IN16B for confined water in the 3 matrices (a) MCM-41, (b) DVA-PMO, and (c) DVB-PMO at three temperatures 278 K, 258 K, and 243 K (filled symbols). Fits using the jump diffusion model (dashed lines).

**Table 3.** Comparison of the translation diffusion coefficient $D_T$ and the residence time $\tau_0$ of water molecules derived from the jump diffusion model for two different experimental resolutions.

| **Temperature (K)** | **$D_t$ ($10^{-9}$ $m^2.s^{-1}$)** **$\tau_0$ (ps)** | **MCM-41** | | **DVA-PMO** | | **DVB-PMO** | |
|---|---|---|---|---|---|---|---|
| | | **IN5B (30 ps)** | **IN16B (1 ns)** | **IN5B (30 ps)** | **IN16B (1 ns)** | **IN5B (30 ps)** | **IN16B (1 ns)** |
| **278** | $D_t$ | 1.14 | 0.78 | 1.20 | 0.72 | 1.30 | 1.03 |
| | $\tau_0$ | 4.5 | 22 | 6.1 | 27 | 4.6 | 23 |
| **258** | $D_t$ | 0.56 | 0.34 | 0.71 | 0.33 | 0.64 | 0.52 |
| | $\tau_0$ | 12.0 | 26 | 12.4 | 31 | 11.0 | 27 |
| **243** | $D_t$ | 0.26 | 0.13 | 0.57 | 0.13 | 0.32 | 0.17 |
| | $\tau_0$ | 23.7 | 37 | 28.6 | 44 | 33.9 | 35 |

The linewidth of the quasielastic intensity measured on IN16B is illustrated in Fig. 8. It is proportional to the squared momentum transfer in the zero-$Q$ limit, and gradually deviates from the Fickian diffusion regime at larger $Q$, as also seen on IN5B. In some cases, more markedly for water filled DVB-PMO at 278 K, one observed variations of the measured linewidth with respect to a smooth monotonous increase. They are attributed to statistical fluctuations, which are more visible at the highest temperature when the quasielastic broadening approaches the maximum energy transfer on IN16B (30 μeV corresponding to 0.045 $ps^{-1}$), while the quasielastic intensity increases and the statistical fluctuations vanish as the line sharpens at lower temperature. As for IN5B, the linewidth was fitted with a jump-diffusion model. The diffusion coefficients and residences times derived from these fits are illustrated in Fig. 9, and also given in Table 3. The diffusion coefficient and the residence time of bulk water are added in Fig. 9, which recalls the presentation made in Fig. 7 for IN5B results. The diffusion coefficients measured on IN16B are systematically smaller (by a factor of 1.2 to 2) than those measured on IN5B for the same water filled matrices. Also, the residence times are longer. This confirms that the fraction of the water molecules, which was denoted as non-mobile on IN5B, comprises particles that actually diffuse but much more slowly. We inferred from IN5B results that the molecules performing 'bulk-like' spatially restricted motions, were located in the center of the pore. Accordingly, we relate the sharpest quasielastic contribution observed on IN16B to a slower diffusion process, which corresponds to molecules interacting with the interface of water filled matrices. Importantly, the diffusivity of these interfacial water molecules depends on the matrix chemistry. The largest reduction of translational mobility is obtained for the two hydrophilic matrices (MCM-41 and DVA-PMO). For purely siliceous MCM-41, this slowdown of the diffusion of interfacial water molecules is classically attributed to the formation of H-bonds between water and silanols. For DVA-PMO, silanol adsorption sites are dispersed on the pore surface due to the bridging organic units that impose a separation distance of 1.2 nm between silica units in the direction parallel to the pore axis. However, this effect is balanced by the presence of amino groups between silica units, which act as secondary H-bonding sites for interfacial water molecules. This is supported by the observation by NMR of strong interfacial correlations between water, and both silica and organic units.[33]

For water filled DVB-PMO, intermediate values of the diffusivity that stand between the bulk and the hydrophilic confinement situation are observed. Again, this points to an effect of the dilution of silica units, which are regularly spaced every 1.2 nm along the pore channel axis. Unlike DVA-PMO, this reduction of the number of H-bonding silanol groups is not counterbalanced by the DVB organic spacers that have no polar groups and cannot act as secondary adsorbing sites. This interpretation is in line with the observation of depleted regions in the vicinity of the aprotic organic bridging units.[33]

As a whole, the relation between surface chemistry and interfacial layer mobility, although studied for a limited number of materials, agrees with the evaluation of the relative surface hydrophilicity made for series of PMOs and quantified in terms water capillary pressure and surface water coverage prior to capillary condensation.[36]

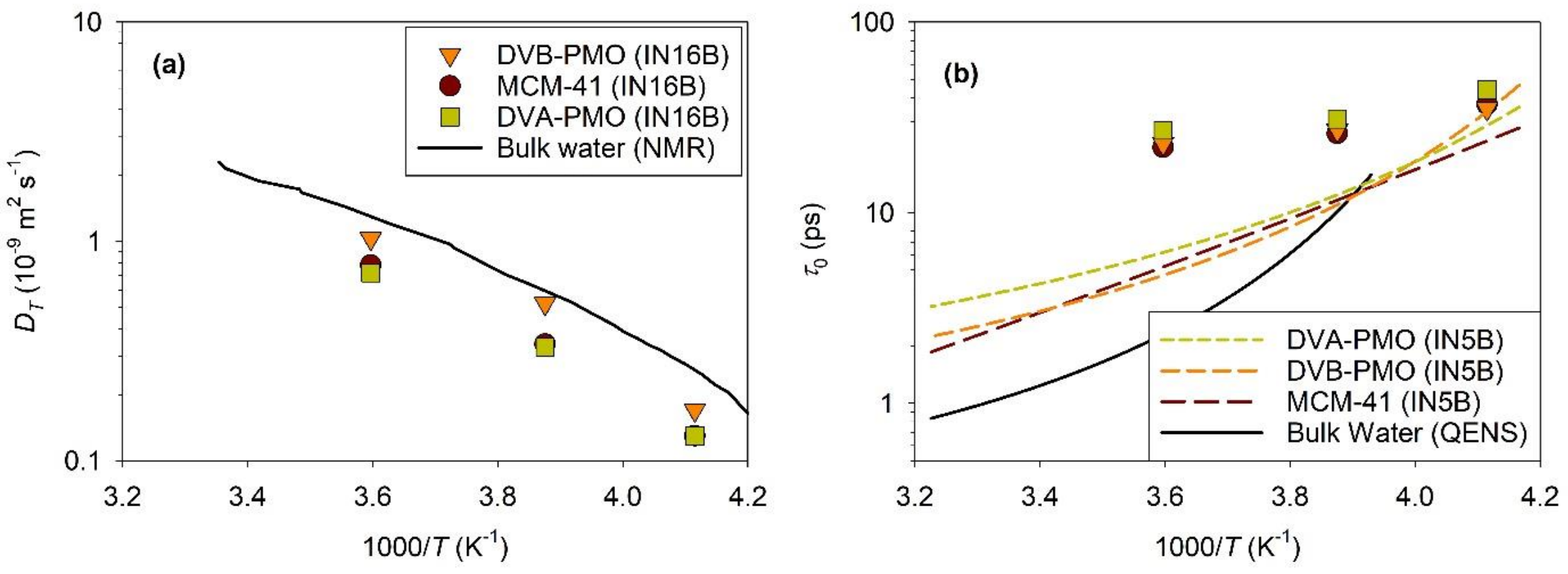

**FIG. 9**. (a) Translational diffusion coefficient $D_T$ and (b) residence time $\tau_0$ of water evaluated from the fit of IN16B spectra with a jump-diffusion model for water confined in MCM-41 and two PMOs as a function of the inverse temperature (symbols). In panel a, the NMR diffusion coefficient of bulk water from Price *et al.*[56] (solid line). In panel b, the residence time of water confined in the same matrices fitted from IN5B spectra (dashed line) and the residence time of bulk water from Teixeira *et al.*[50] (solid line)

Some important differences can be underlined from the comparison of the diffusivities obtained in the present QENS study and those derived from PFG NMR on similar samples.[34] A greater reduction of the diffusivity with respect to bulk water was obtained in the latter case, as well as a larger dependence on the nature of the porous sample (by up to two orders of magnitude). The apparent disagreement between the diffusivities measured by the two methods is intriguing. As mentioned in the introduction part, the diffusivities measured by both methods are averaged on very different molecular displacements (i.e. ~ 1 Å for QENS and ~ 1 μm for PFG NMR). The latter displacement exceeds notably the size of each individual porous grain. As thoroughly explained in this PFG NMR study, the obtained long-range diffusivities result from the migration of water molecules in the pores, but also on the outer surface of the porous grains and in the interparticle space. The transport behavior depends on the rate exchange between these different phases and their physical states (liquid or gaseous). Under the same filling conditions as those used in the QENS study (i.e. absence of liquid excess) and for loosely agglomerated particles (such as MCM-41), it was concluded that the average diffusivity $D_{av}$ was barely related to the liquid phase $D_l$ but dominated by properties of the vapor phase with $D_{av}=p_vD_v$, where $p_v<<1$ and $D_v>>D_l$ are the relative fraction and the diffusivity of water in the vapor phase. On the contrary, for agglomerated particles (such as DVA-PMO), or when an excess of bulk water has crystallized in the interparticle space, it was concluded that the long range transport of water molecules was occurring in a liquid thin film covering the external surface on the grains. In this case, the measured slow diffusive component was shown to be an effective value that depends on the macroscopic rather than nanoporous structure of the sample in terms of particle morphology and their agglomeration.

On the contrary, the short-range diffusivities measured by QENS probe the transport of water molecules inside a single pore. As such, they are intrinsic properties of the confined liquid phase, which were shown to be affected by the pore size, and the surface interaction. However, they bring no information on the macroscopic transport mechanisms of water across the porous materials, which is important in many natural processes and for technological applications where fluid flow and imbibition occur. In this sense, we emphasize that the combination of PFG NMR and QENS methods to study the dynamics of confined water in PMOs provides crucial information on the two sides of the same coin.

## CONCLUSION

We performed a systematic QENS study of the dynamics of liquid water confined in mesoporous silica (MCM-41) and organosilicas (PMOs) with different surface chemistries in a temperature range spanning from 243 to 300 K. Two high-flux spectrometers with complementary energy resolution were combined so as to cover an extended dynamical range. Following recent recommendations made for bulk water, a model free analysis of the incoherent dynamic structure factor was applied for the first time to confined water. The analysis of IN5B results points towards the existence of two distinct dynamics: a fast local motion (typical time varying from 1 to 3 ps), which implies all the water molecules and translational diffusion dynamics attributable only to a fraction of molecules. The EISF of the former component is consistent with the isotropic rotational diffusion of the water molecule at 300 K. However, this interpretation is questioned at lower temperature. Based on the concept of intra-basin local relaxation introduced for bulk water, our result suggests a reduction of the spatial amplitude of this local motion on cooling. The second quasielastic relaxation is attributed to translational diffusion. It concerns only a fraction of mobile molecules, which are most probably located at the pore center. Deviations from the Fick's law at large $Q$s were accounted by the jump-diffusion model and at low $Q$s by transient spatial restriction effects. The obtained values of the diffusion coefficient $D_T$ and the residence time $\tau_0$ indicate a ‘bulk-like’ behavior. The slowing down of translational diffusion due to confinement is moderate. It is within 10% for $D_T$ but it is more marked for $\tau_0$ that is up to four times longer than for bulk water. The EISFs demonstrate that the translational diffusion is spatially restricted on the instrumental timescale, which differs from bulk water. The radius $R_T$ of the sphere confining the translational diffusion of the mobile molecules decreases from 9 to 5 Å, during cooling from 300 to 243 K. It is not strictly related to the pore size itself. We consider that it rather emerges from the hindrance due to the surrounding molecules that are influenced by the pore interface. This is supported by the increase on cooling of the fraction $p$ of molecules that are dynamically frozen (i.e. not diffusing on the timescale of IN5B), and that are most likely located at the interface. It is about 10-20% at 300 K, and reaches values as large as 40-50% at 243 K, which corresponds to a thickness of the ‘non-mobile’ layer in the range of 1-5 Å. This fraction also varies as a function of the nature of the matrix, illustrating the effect of the surface interaction on the dynamics of the interfacial layer. Finally, our study using the high-resolution spectrometer (IN16B) demonstrates that interfacial molecules are dynamically active on a longer timescale (about 1 ns). Their motion could be described by a jump-diffusion model as well. However, the corresponding characteristic parameters $D_T$ and $\tau_0$ demonstrate a significantly slower process than that of the liquid located in the pore center. Moreover, the diffusivity of interfacial water molecules shows dependence on the nature of the confining matrix. The largest slowdown is obtained for hydrophilic matrices (MCM-41 and DVA-PMO) with respect to the more hydrophobic one (DVB-PMO). As a whole, this study indicates that, on the ps-to-ns timescale and for a pore size of about 3.5 nm, the details of the water/surface interaction (*i.e.* matrix hydrophilicity, and presence of silanol or amino H-bonding sites) hardly affect the generic confinement effects that involve the water molecules located in the core of the pore but they are important in determining the translational motion of the water molecules located at the interface.

## ACKNOWLEDGEMENTS

This work was conducted in the frame of the DFG-ANR collaborative project (Project NanoLiquids N° ANR-18-CE92-0011-01, DFG: FR 1372/25-1 - Projektnummer 407319385, and DFG Hu850/11-1 - Projektnummer 407319385), which is expressly acknowledged. It is a pleasure to acknowledge the Institut Laue-Langevin for the allocation of neutron beam time.[43, 44] This work is part of the Ph.D. thesis of A.J. who benefits from a grant from the French Ministry of Higher Education, Research, and Innovation.

## DATA AVAILABILITY

Raw data were generated at the Institut Laue-Langevin (ILL) large scale facility. They will be available in the ILL Data Portal at doi:10.5291/ILL-DATA.6-07-34 and doi:10.5291/ILL-DATA.6-07-45 following an embargo period.[43, 44] Derived data supporting the findings of this study are available from the corresponding author upon reasonable request.

## SUPPLEMENTARY MATERIAL

Sample Filling: relation between RH and the water mass fraction

Sample Filling: determination of saturated samples and overfilled samples by DSC

Temperature dependence of the QENS spectra measured on IN5B at $Q = 0.8$ Å$^{-1}$ and on IN16B at $Q = 0.85$ Å$^{-1}$ for water filled DVA-PMO and in BP-PMO.

$Q$-dependence of the QENS spectra measured on IN5B at the temperature $T = 278$ K for water filled DVA-PMO and BP-PMO.

Fitting procedure of QENS spectra

Related references

## REFERENCES

1. J. M. Zanotti, G. Gibrat and M. C. Bellissent-Funel, Physical Chemistry Chemical Physics **10** (32), 4865-4870 (2008).
2. B. Bagchi, Chemical Reviews **105** (9), 3197-3219 (2005).
3. L. Garnier, A. Szymczyk, P. Malfreyt and A. Ghoufi, Journal of Physical Chemistry Letters **7** (17), 3371-3376 (2016).
4. Q. Berrod, S. Lyonnard, A. Guillermo, J. Ollivier, B. Frick, A. Manseri, B. Ameduri and G. Gebel, Macromolecules **48** (17), 6166-6176 (2015).
5. N. Martinez, A. Morin, Q. Berrod, B. Frick, J. Ivier, L. Porcar, G. Gebel and S. Lyonnard, Journal of Physical Chemistry C **122** (2), 1103-1108 (2018).
6. L. Bocquet, Nature Materials **19** (3), 254-256 (2020).
7. J. Swenson, R. Bergman and W. S. Howells, Journal of Chemical Physics **113** (7), 2873-2879 (2000).

8. N. Malikova, S. Longeville, J. M. Zanotti, E. Dubois, V. Marry, P. Turq and J. Ollivier, Physical Review Letters **101** (26), 265901 (2008).

9. V. Marry, E. Dubois, N. Malikova, J. Breu and W. Haussler, Journal of Physical Chemistry C **117** (29), 15106-15115 (2013).

10. S. Cerveny, F. Mallamace, J. Swenson, M. Vogel and L. M. Xu, Chemical Reviews **116** (13), 7608-7625 (2016).

11. D. Demuth, M. Sattig, E. Steinrucken, M. Weigler and M. Vogel, Zeitschrift Fur Physikalische Chemie-International Journal of Research in Physical Chemistry & Chemical Physics **232** (7-8), 1059-1087 (2018).

12. C. Lederle, M. Sattig and M. Vogel, Journal of Physical Chemistry C **122** (27), 15427-15434 (2018).

13. M. Weigler, M. Brodrecht, G. Buntkowsky and M. Vogel, Journal of Physical Chemistry B **123** (9), 2123-2134 (2019).

14. J. Swenson and S. Cerveny, Journal of Physics-Condensed Matter **27** (3) (2015).

15. A. A. Milischuk and B. M. Ladanyi, Journal of Chemical Physics **135** (17) (2011).

16. P. Gallo, M. Rovere and S. H. Chen, Journal of Physics-Condensed Matter **24** (6) (2012).

17. N. Kuon, A. A. Milischuk, B. M. Ladanyi and E. Flenner, Journal of Chemical Physics **146** (21) (2017).

18. M. C. Bellissent-Funel, S. H. Chen and J. M. Zanotti, Physical Review E **51** (5), 4558-4569 (1995).

19. J. M. Zanotti, M. C. Bellissent-Funel and S. H. Chen, Physical Review E **59** (3), 3084-3093 (1999).

20. S. Takahara, M. Nakano, S. Kittaka, Y. Kuroda, T. Mori, H. Hamano and T. Yamaguchi, Journal of Physical Chemistry B **103** (28), 5814-5819 (1999).

21. A. Faraone, L. Liu, C. Y. Mou, P. C. Shih, J. R. D. Copley and S. H. Chen, Journal of Chemical Physics **119** (7), 3963-3971 (2003).

22. S. Takahara, N. Sumiyama, S. Kittaka, T. Yamaguchi and M. C. Bellissent-Funel, Journal of Physical Chemistry B **109** (22), 11231-11239 (2005).

23. A. Faraone, K. H. Liu, C. Y. Mou, Y. Zhang and S. H. Chen, Journal of Chemical Physics **130** (13) (2009).

24. I. M. Briman, D. Rebiscoul, O. Diat, J. M. Zanotti, P. Jollivet, P. Barboux and S. Gin, Journal of Physical Chemistry C **116** (12), 7021-7028 (2012).

25. M. Aso, K. Ito, H. Sugino, K. Yoshida, T. Yamada, O. Yamamuro, S. Inagaki and T. Yamaguchi, Pure and Applied Chemistry **85** (1), 289-305 (2013).

26. M. Baum, F. Rieutord, F. Jurany, C. Rey and D. Rebiscoul, Langmuir **35** (33), 10780-10794 (2019).

27. S. O. Diallo, L. Vlcek, E. Mamontov, J. K. Keum, J. H. Chen, J. S. Hayes and A. A. Chialvo, Physical Review E **91** (2) (2015).

28. M. C. Bellissent-Funel, K. Kaneko, T. Ohba, M. S. Appavou, A. J. Soininen and J. Wuttke, Physical Review E **93** (2) (2016).

29. A. Kiwilsza, A. Pajzderska, M. A. Gonzalez, J. Mielcarek and J. Wasicki, Journal of Physical Chemistry C **119** (29), 16578-16586 (2015).

30. J. Beck, J. Vartuli, W. Roth, M. Leonowicz, C. Kresge, K. Schmitt, C. Chu, D. Olson, E. Sheppard, S. Mccullen, J. Higgins and J. Schlenker, Journal of the American Chemical Society **114** (27), 10834-10843 (1992).

31. M. Grun, I. Lauer and K. K. Unger, Advanced Materials **9** (3), 254-& (1997).

32. F. Hoffmann, M. Cornelius, J. Morell and M. Fröba, Angewandte Chemie-International Edition **45** (20), 3216-3251 (2006).

33. J. B. Mietner, F. J. Brieler, Y. J. Lee and M. Fröba, Angewandte Chemie-International Edition **56** (40), 12348-12351 (2017).

34. J. B. Mietner, M. Fröba and R. Valiullin, Journal of Physical Chemistry C **122** (24), 12673-12680 (2018).

35. M. Thommes, J. Morell, K. A. Cychosz and M. Fröba, Langmuir **29** (48), 14893-14902 (2013).

36. J. B. Mietner, Visiting Nanopores : the great potential of PMOs for studying the properties of water in nanopores of different polarity, Doctoral dissertation, Universität Hamburg, 2018.

37. J. Qvist, H. Schober and B. Halle, Journal of Chemical Physics **134** (14) (2011).

38. D. Morineau, Y. D. Xia and C. Alba-Simionesco, Journal of Chemical Physics **117** (19), 8966-8972 (2002).

39. D. Morineau and C. Alba-Simionesco, Journal of Chemical Physics **118** (20), 9389-9400 (2003).

40. D. Morineau and C. Alba-Simionesco, Journal of Physical Chemistry Letters **1** (7), 1155-1159 (2010).

41. A. R. A. Hamid, R. Mhanna, P. Catrou, Y. Bulteau, R. Lefort and D. Morineau, Journal of Physical Chemistry C **120** (20), 11049-11053 (2016).

42. A. Hamid, R. Mhanna, R. Lefort, A. Ghoufi, C. Alba-Simionesco, B. Frick and D. Morineau, Journal of Physical Chemistry C **120** (17), 9245-9252 (2016).

43. D. Morineau, M. Busch, M. Fröba, P. Huber and J. Ollivier, in *Water dynamics in mesoporous with periodically alternating surface interaction. Institut Laue-Langevin (ILL) ;* (doi: 10.5291/ILL-DATA.6-07-34 (2018)).

44. D. Morineau, M. Appel, M. Busch, B. Frick, M. Fröba, P. Huber, A. Jani and J. Ollivier, in *Water dynamics in mesoporous with periodically alternating surface interaction. Institut Laue-Langevin (ILL) ;* (doi:10.5291/ILL-DATA.6-07-45 (2019)).

45. M. Appel and B. Frick, Review of Scientific Instruments **88** (3) (2017).

46. O. Arnold, J. C. Bilheux, J. M. Borreguero, A. Buts, S. I. Campbell, L. Chapon, M. Doucet, N. Draper, R. F. Leal, M. A. Gigg, V. E. Lynch, A. Markvardsen, D. J. Mikkelson, R. L. Mikkelson, R. Miller, K. Palmen, P. Parker, G. Passos, T. G. Perring, P. F. Peterson, S. Ren, M. A. Reuter, A. T. Savici, J. W. Taylor, R. J. Taylor, R. Tolchenoy, W. Zhou and J. Zikoysky, Nuclear Instruments & Methods in Physics Research Section a-Accelerators Spectrometers Detectors and Associated Equipment **764**, 156-166 (2014).

47. D. Richard, M. Ferrand and G. J. Kearley, Journal of Neutron Research **4**, 33-39 (1996).

48. A. Arbe, P. M. de Molina, F. Alvarez, B. Frick and J. Colmenero, Physical Review Letters **117** (18) (2016).

49. M. Bée, *Quasi-elastic Neutron Scattering Principles and Application in Solid State Chemistry, Biology and Materials Science.*, Adam Hilger, Bristol ed. (1998).

50. J. Teixeira, M. C. Bellissentfunel, S. H. Chen and A. J. Dianoux, Physical Review A **31** (3), 1913-1917 (1985).

51. F. Volino and A. J. Dianoux, Molecular Physics **41** (2), 271-279 (1980).

52. G. H. Findenegg, S. Jahnert, D. Akcakayiran and A. Schreiber, ChemPhysChem **9** (18), 2651-2659 (2008).

53. S. Jahnert, F. V. Chavez, G. E. Schaumann, A. Schreiber, M. Schonhoff and G. H. Findenegg, Physical Chemistry Chemical Physics **10** (39), 6039-6051 (2008).

54. S. Gruener, T. Hofmann, D. Wallacher, A. V. Kityk and P. Huber, Physical Review E **79** (6) (2009).

55. P. L. Hall and D. K. Ross, Molecular Physics **42** (3), 673-682 (1981).

56. W. S. Price, H. Ide and Y. Arata, Journal of Physical Chemistry A **103** (4), 448-450 (1999).

57. M. D. Ediger and P. Harrowell, Journal of Chemical Physics **137** (8) (2012).

58. R. Richert, Annual Review of Physical Chemistry, Vol 62 **62**, 65-84 (2011).

# Dynamics of Water Confined in Mesopores with Variable Surface Interaction

Aîcha Jani[a], Mark Busch[b], J. Benedikt Mietner[c], Jacques Ollivier[d], Markus Appel[d], Bernhard Frick[d], Jean-Marc Zanotti[e], Aziz Ghoufi[a], Patrick Huber[b,f,g*], Michael Fröba[c*], and Denis Morineau[a*]

[a]Institute of Physics of Rennes, CNRS-University of Rennes 1, UMR 6251, F-35042 Rennes, France

[b]Hamburg University of Technology, Center for Integrated Multiscale Materials Systems CIMMS, 21073 Hamburg, Germany

[c]Institute of Inorganic and Applied Chemistry, University of Hamburg, 20146 Hamburg, Germany

[d]Institut Laue-Langevin, 71 avenue des Martyrs, F-38000 Grenoble, France

[e]Laboratoire Léon Brillouin, CEA, CNRS, Université Paris-Saclay, F-91191 Gif-sur-Yvette, France

[f]Deutsches Elektronen-Synchrotron DESY, Centre for X-Ray and Nano Science CXNS, 22603 Hamburg, Germany

[g]Hamburg University, Centre for Hybrid Nanostructures CHyN, 22607 Hamburg, Germany

* denis.morineau@univ-rennes1.fr, froeba@chemie.uni-hamburg.de, patrick.huber@tuhh.de

# Supporting Information

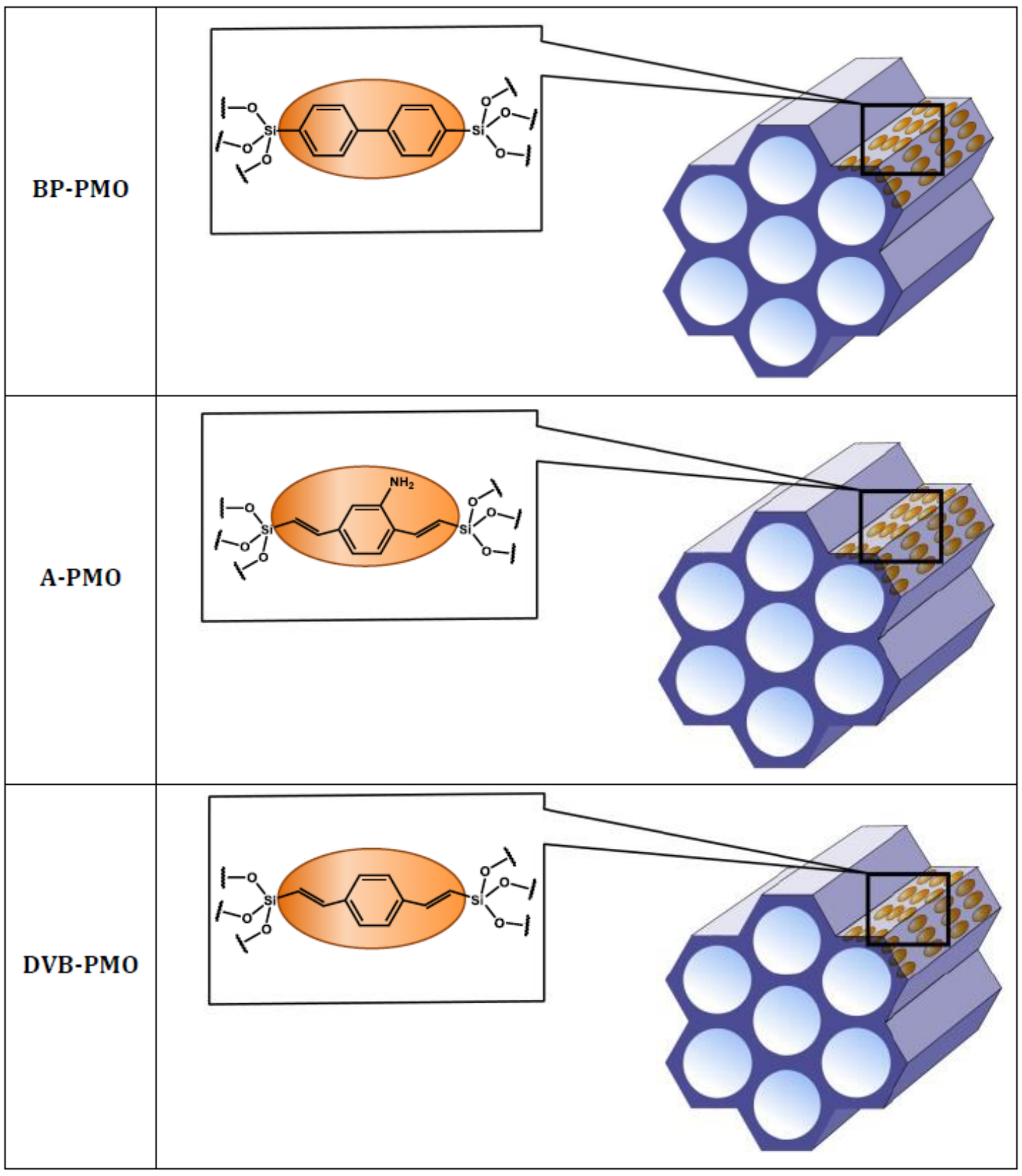

**Fig. S1:** Schematics of the three PMOs studied. From top to bottom: biphenyl-bridged PMO, divinyl-aniline-bridged PMO, and divinyl-benzene-bridged PMO.

**Sample Filling: relation between RH and the water mass fraction**

The filling procedure that was applied in this study can be considered as an experimental realization of the grand canonical thermodynamic ensemble, where the water chemical potential of the gas phase is imposed by the relative humidity (i.e. RH = $P_{water}/P_{sat}$). At given temperature, the relation between the amount of water that fills each porous medium and the relative pressure of the saturating atmosphere is determined by its water-vapor adsorption isotherm. They were measured for series of PMOs that were similar to those used in the present study.[1, 2] They were all characteristic of type V isotherms: at low relative pressure, the isotherm exhibits a low adsorption region, followed by a pore capillary condensation step at an intermediate pressure (0.5-0.6), and reaches a plateau where the amount of the water increases only slightly with the increase of pressure.

The relative humidity (RH=75%) used in the present study imposes a partial pressure of water that is properly situated between the capillary condensation pressure and the saturation pressure. Under this condition, the amount of confined water is defined by the level of the plateau region of the adsorption isotherm, which also corresponds to the filling of the porosity with condensed liquid without bulk excess liquid.

When possible, it is noteworthy that it is preferable to impose the relative pressure rather than the mass fraction of water that fills the porous materials. Incidentally, the former is easily achieved for water through the control of RH. It insures that all the different water-filled materials result from the equilibrium with water vapor at the same chemical potential, which makes their comparison more sensible from the thermodynamic point of view. Moreover, from a practical point of view, it is more difficult to finely adjust the water amount by simple addition of a given mass or volume of liquid that conforms to the level of the plateau region of the isotherm.

We would also like to comment that selecting different hydration levels that would result in the same water amount per mg of material is inappropriate. It does not guarantee that the different systems are in the same physical state. Indeed, the different PMOs have different porous volume (per mg of material), because their structures differ, but also because the mass density of their walls is affected by the replacement of silica units by lighter atoms forming the organic bridges. In general, applying this filling approach to a series of PMOs requires that some of them must be overfilled with bulk excess (i.e. beyond the saturation pressure) or must be underfilled (i.e. below the capillary pressure). In both cases, they cannot be compared with the saturated samples. This is obvious in Fig. S2a, where the relative amount of water is illustrated at RH = 75% (saturated samples) and RH =100% (overfilled samples) for the four materials, expressed in terms of mass of confined water per unit mass of porous matrix.

The determination of the relative amount of water that is achieved under the conditions used for QENS samples is limited by the large contribution of the aluminum cell to the total sample mass. In this case, the relative mass of water ranges between 0.1% and 0.2%. It was determined with better accuracy (gaining a factor of about 20) for a series of companion samples that were enclosed in lighter DSC pans and submitted to the same filling conditions. The corresponding compositions are provided in Table S1. The filling fraction were calculated in two different ways: $x$ is the amount of water $m_w$ per mass of matrix $m_p$, ($x = m_w/m_p$) and $\rho$ is the amount of water divided by the matrix pore volume ($\rho = x/ V_p$), with $V_p$ being determined by nitrogen adsorption isotherms. These results are also illustrated in Fig. S2. Notably, for the saturated samples (RH = 75%) the filling fraction $x$ varies much among the different porous materials (up to a factor of 3), which is partly due to different porous volumes. Indeed, more similar values are obtained when the filling fraction is scaled to the porous volume, although differences persist. As discussed above, this illegitimates any attempt to compare samples that would have been filled at different hydration levels so as to match their mass fraction $x$.

**Table S1:** Composition of the water filled samples equilibrated at two different RH (saturated samples RH = 75%, and overfilled samples RH = 100%).

| **Porous Matrix** | **Pore Volume $V_p$ [a] (cm³/g)** | **RH** | **Matrix mass $m_p$ (mg)** | **Water mass $m_w$ (mg)** | **Filling fraction $x$ [b] (g/g)** | **Filling fraction $\rho$ [c] (g/cm³)** |
|---|---|---|---|---|---|---|
| **MCM-41** | 0.893 | 75% | 4.28 | 2.52 | 0.659 | 0.589 |
| | | 100% | 3.81 | 2.52 | 0.661 | 0.740 |
| **DVA-PMO** | 0.890 | 75% | 3.75 | 1.66 | 0.443 | 0.498 |
| | | 100% | 3.52 | 1.83 | 0.520 | 0.584 |
| **BP-PMO** | 0.445 | 75% | 5.27 | 1.11 | 0.211 | 0.474 |
| | | 100% | 3.61 | 0.94 | 0.260 | 0.584 |
| **DVB-PMO** | 0.990 | 75% | 1.13 | 0.47 | 0.420 | 0.424 |
| | | 100% | 1.64 | 1.31 | 0.807 | 0.99 |

[a]: evaluated from the nitrogen physisorption isotherms. [b]: defined as the amount of water per mass of matrix, [c]: defined as the amount of water per pore volume.

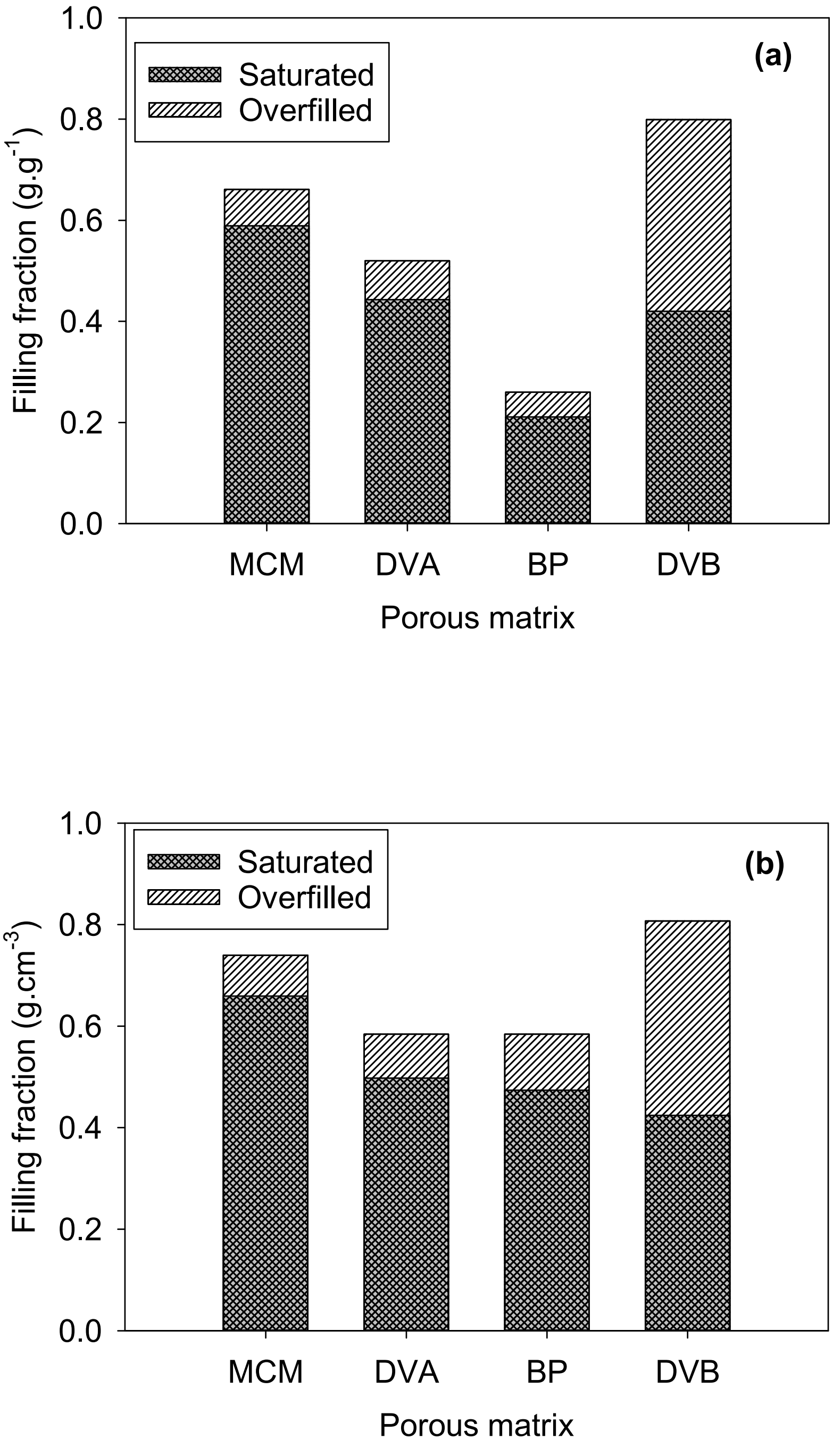

**Fig. S2:** (a) Water filling fraction $x$ expressed in terms of mass of confined water per unit mass of porous matrix. (b) Water filling fraction $\rho$ expressed in terms of mass of confined water per unit porous volume, as determined by nitrogen physisorption. Capillary filled and overfilled samples were obtained after equilibration in a chamber containing a saturated NaCl solution (RH = 75%), and pure water (RH = 100%), respectively.

**Sample Filling: determination of saturated samples and overfilled samples by DSC**

Differential scanning calorimetry (DSC) experiments were performed with a Q-20 TA Instrument equipped with a liquid nitrogen cooling system and using Tzero hermetic aluminum pans. The melting transition of an indium sample was used for calibration of temperature and heat flux. Temperature scanning rates of 10 $K \cdot min^{-1}$ were applied during an initial cooling ramp and a subsequent heating ramp. The composition of the samples that were water filled at two different relative pressure (saturated samples RH = 75%, and overfilled samples RH = 100%) is described in details in the previous part (cf. Table S1).

For all the four matrices, the porous materials filled at RH = 75% presented an exothermic peak, with an onset located in the range 227-237 K, which indicated the freezing of confined water. This onset of crystallization is also close to the temperature of the maximum position of the endothermic peak observed on heating that is due the melting of nanoconfined ice. The absence of thermal event around 273 K demonstrates unambiguously that these capillary filled samples do not contain bulk excess liquid outside the porosity.

On the contrary, samples filled at RH = 100% presented an additional thermal event around 273 K, due to excess water that is presumably located around the grains of the porous powder as well as in interstitial regions and behaves more like bulk water. This demonstrates that these samples were overfilled. For these samples, a quantification of the relative amount of confined and bulk liquid, based on the relative intensity of the two peaks is hardly possible because the melting enthalpy of ice is reduced by confinement. However, fruitful conclusions can be achieved based on their relative masses (see Fig. S2 and Table S1 and earlier discussion). Except for DVB that obviously contained a larger amount of excess water, as seen by the prominent thermal event at 273 K, the difference of mass fraction between samples filled at RH = 75% and RH = 100% is small. In full agreement with the water adsorption isotherms, this demonstrates that the filling of the porosity with water had essentially reached its saturating limit at RH = 75%.

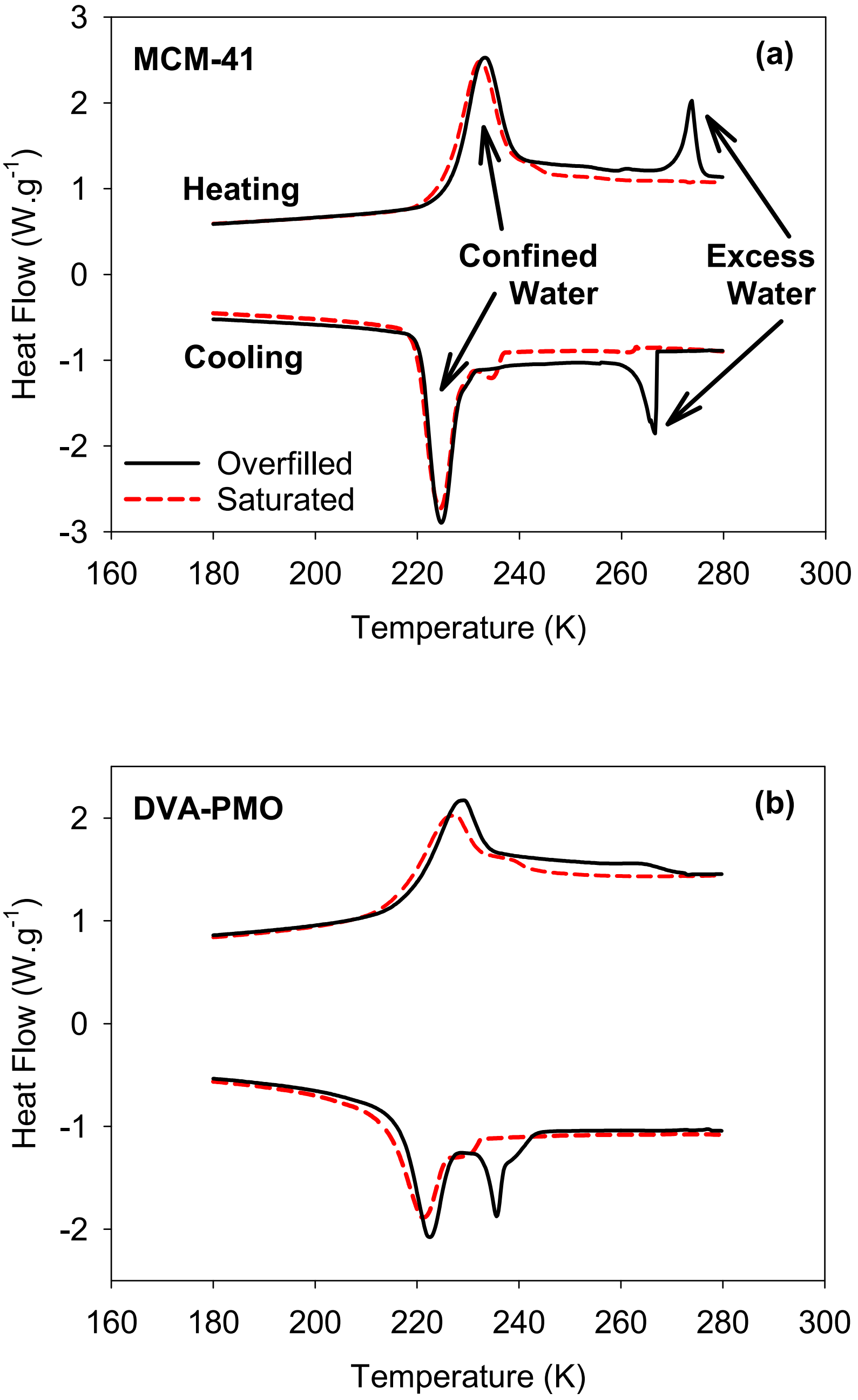

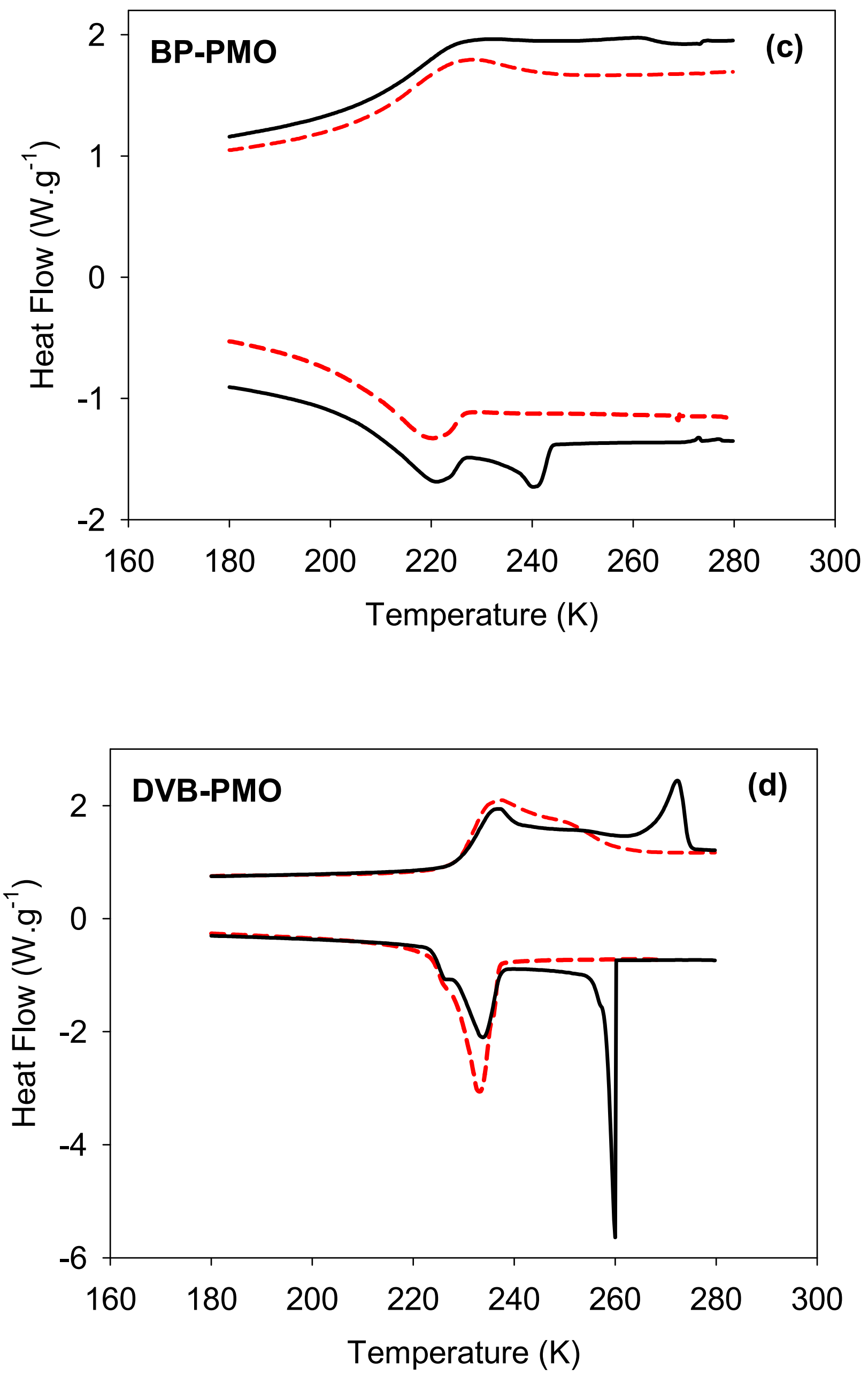

**Fig. S3:** DSC scans (endotherm up) of water filled (a) MCM-41, (b) DVA-PMO, (c) BP-PMO, and (d) DVB-PMO. The thermograms of capillary filled (solid line) and overfilled samples (dashed line) were recorded on cooling (lower curves) and on heating (upper curves) with a temperature scanning rate of 10 K.min$^{-1}$.

## QENS spectra of water filled samples

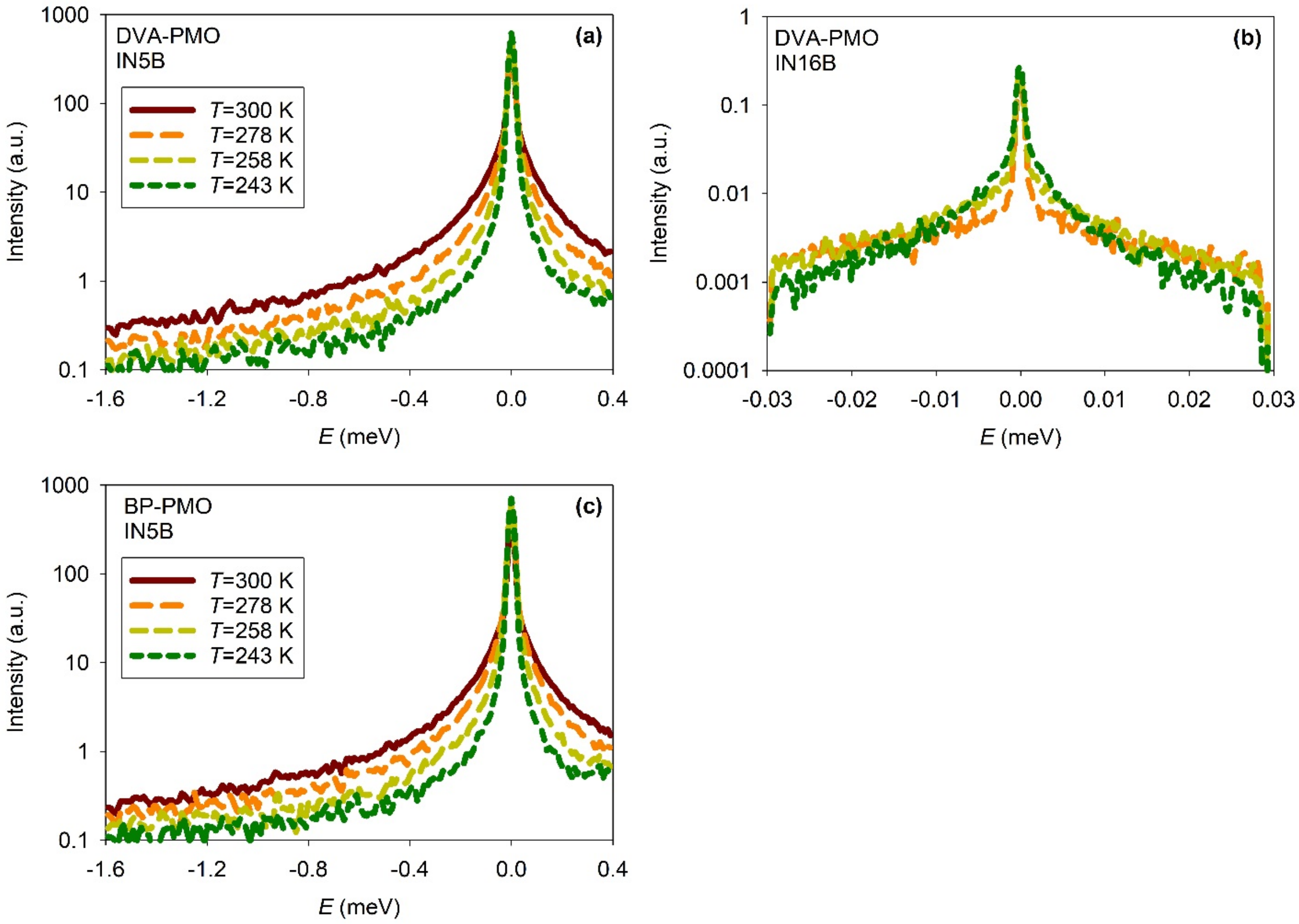

**FIG. S4.** Temperature dependence of the scattering intensity of water filled DVA-PMO (upper panels a, b) and BP-PMO (c). QENS spectra measured on IN5B at $Q = 0.8$ Å$^{-1}$ (left panels a, c) and on IN16B at $Q = 0.85$ Å$^{-1}$ (right panel b).

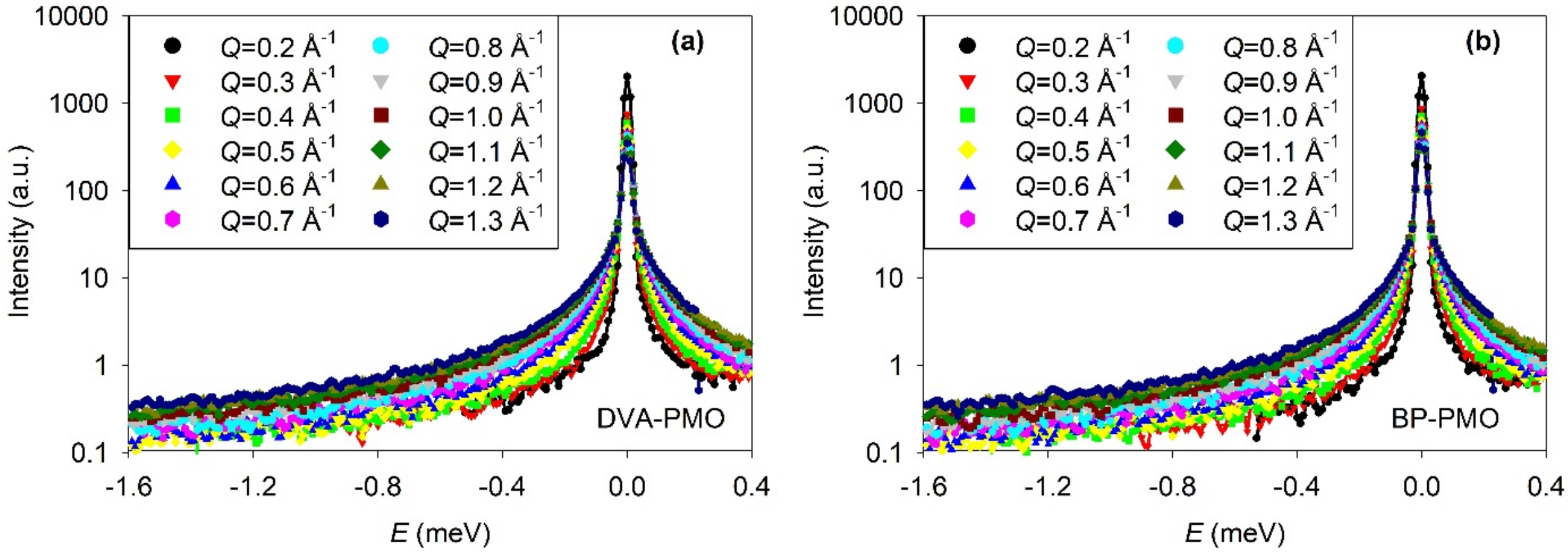

**FIG. S5.** *Q*-dependence of the scattering intensity measured on IN5B at the temperature $T = 278$ K of water filled DVA-PMO (left panel a) and BP-PMO (right panel b).

**Fitting procedure**

All spectra were fitted individually at each $Q$ with a model comprising one elastic component, i.e. a Dirac function $\delta(\omega)$, and a quasielastic contribution, which was approximated by one (IN16B) or a sum of two (IN5B) Lorentzian functions. Importantly, we did not assume any specific $Q$-dependence of the parameters, and therefore no implicit reference to a theoretical model was made. This reduced the possible introduction of bias in the data analysis, as demonstrated in the recent QENS study of bulk water.[3] The essentially elastic contribution arising from the pore wall atoms was accounted for by adding the scattered experimental intensity of the empty matrices to the theoretical functions, which were fitted to the experimental spectra of the water filled matrices. This means that the fitted Dirac function can be actually ascribed to an elastic contribution of water molecules. For the two spectrometers, the scattering intensity of the empty matrices essentially presented an elastic peak and a small background. A tiny dependence on the temperature was observed on IN5B, with the elastic intensity varying by about 4-8% on the temperature range covered by this study (243-300K). This variation is likely due to a Debye Waller factor effect. This is in agreement with the absence of measurable quasielastic intensity and suggests that the dynamics of the framework of the porous matrices is dominated by vibrational modes. Beside the elastic contribution due to the empty matrices, an additional elastic term $A_0(Q)$ was used to fit scattering intensity of the capillary filled samples. The latter can be ascribed to the elastic contribution of water molecules. On IN5B, two Lorentzian functions were needed to reproduce the quasielastic lineshape attributed to water. On IN16B, one Lorentzian function was sufficient. No additional background, such as a constant term, was used in the fitting procedure. The modelled functions were convoluted by the instrument resolution $R(Q,\omega)$ obtained by measuring a vanadium sample according to:

$$I^{\mathrm{IN5B}}(Q,\omega) = [A_0(Q)\delta(\omega) + A_1(Q)L_1(Q,\omega,\Gamma_1) + A_2(Q)L_2(Q,\omega,\Gamma_2)]\otimes R(Q,\omega) \qquad (1)$$

$$I^{\mathrm{IN16B}}(Q,\omega) = [A_0(Q)\delta(\omega) + A_1(Q)L_1(Q,\omega,\Gamma_1)]\otimes R(Q,\omega) \tag{2}$$

where $A_i(Q)$ is the intensity of the $i^{\text{th}}$ component and $L_i$ is a Lorentzian function with a linewidth HWHM $\Gamma_i$. The comparison between the fitted functions and the experimental QENS spectra acquired on both spectrometers is illustrated in Fig. S6 for water filled DVB-PMO and MCM-41. It confirms the very good agreement between the fitted function (red solid line) and the experimental data points (symbol), which match within experimental uncertainties. The additional curves shown in Fig. S6 illustrate the decomposition of the total intensity into one elastic and one (IN16B) or two (IN5B) quasielastic Lorentzian functions.

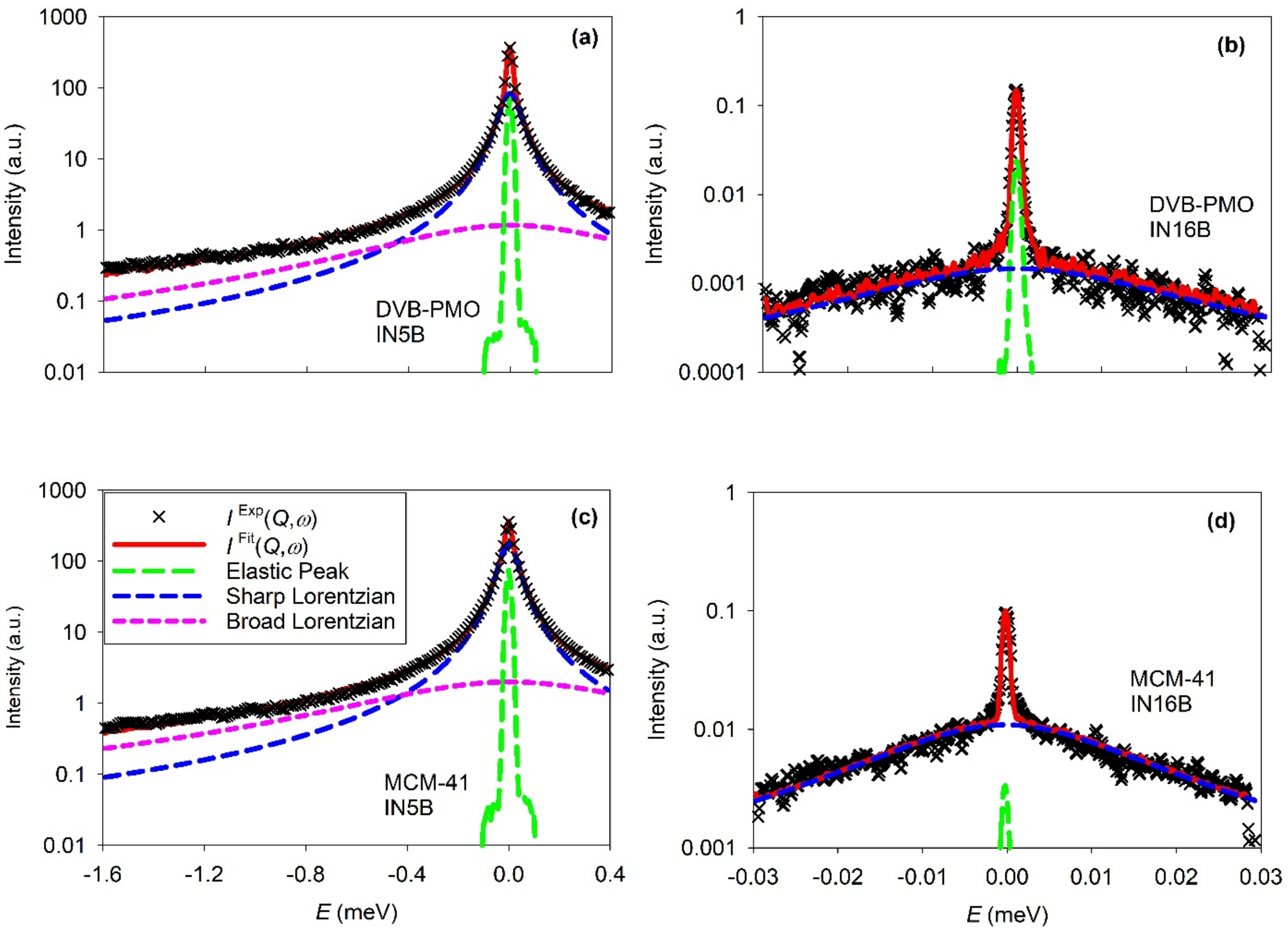

**FIG. S6.** QENS spectra (cross symbols) and fitted functions (lines) at the temperature $T$ = 278 K of water filled DVB-PMO (upper panels a, b) and MCM-41 (lower panels c, d). QENS spectra measured on IN5B at $Q$ = 0.8 Å$^{-1}$ (left panels a, c) and on IN16B at $Q$ = 0.85 Å$^{-1}$ (right panels b, d).

We assumed that the two Lorentzian functions observed on IN5B reflected the presence of two independent dynamics. In the pioneering QENS studies of bulk water, two motions were identified and attributed to continuous rotational diffusion and jump-like translation.[4, 5] QENS studies on bulk [6, 7] and confined water have later adopted this model.[8-10] In some studies, this model was also invoked, but because a simplified version of the fitted functional form was considered, the reference to rotational diffusion was actually lost.[11, 12] The continuous nature of the rotational motion of water molecules has been questioned by experiments and simulations,[13, 14] and it was also argued that the assumptions underlying this QENS model might not be strictly verified.[3] To limit any interpretational bias, we first performed a model-free fit of the data. Both the intensity and the width of the two Lorentzians were determined independently at each $Q$. On IN5B, the first fit revealed that the linewidth of the second (broadest) Lorentzian $\Gamma_2$ was barely varying with $Q$. Contrariwise, the sharpest Lorentzian $\Gamma_1$ was dispersive for both spectrometers. In order to get a more robust fit, $\Gamma_2$ was fixed to its averaged value at each temperature, and the values of all the other parameters were refined again.

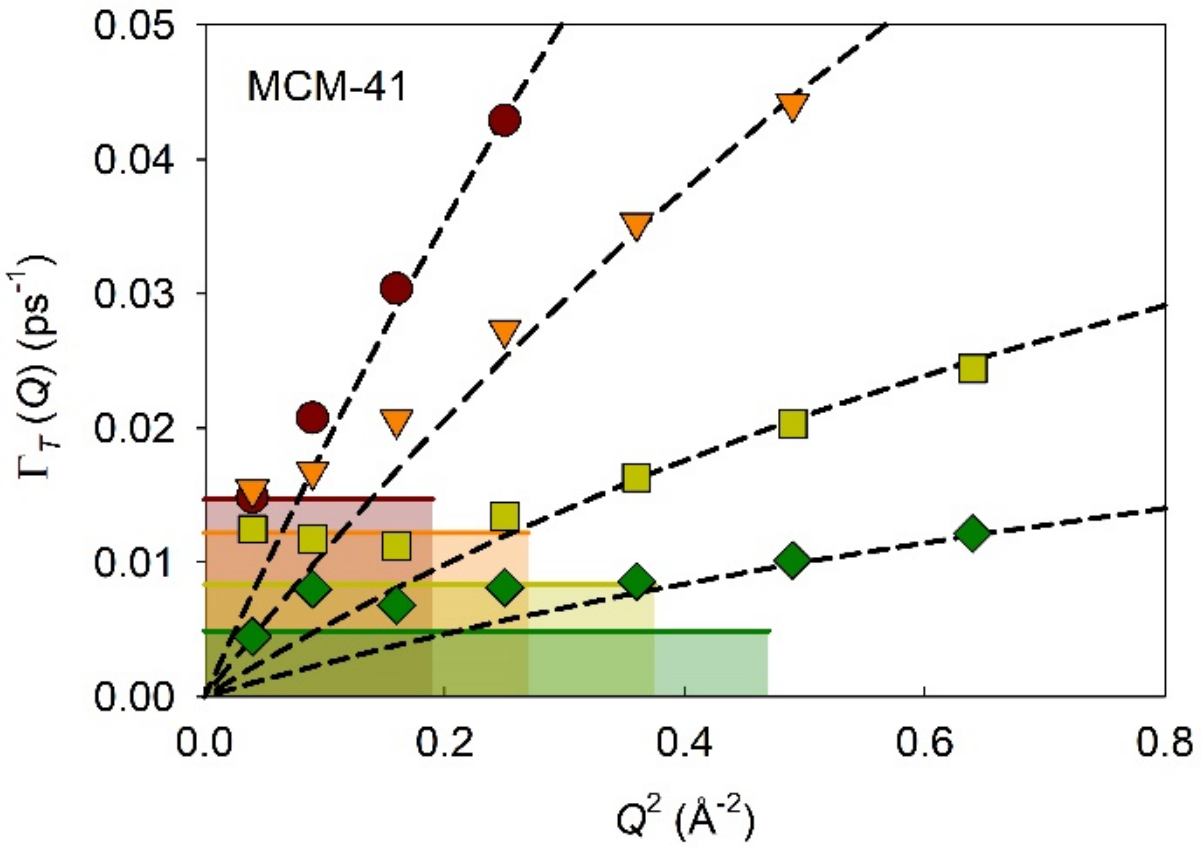

**FIG. S7**. Enlarged view of the half width at half-maximum $\Gamma_T$ in the region of small $Q$s from IN5B for water filled MCM-41 at four temperatures, 300, 278, 258, and 243K, from top to bottom (filled symbols). Fit using the jump diffusion model (thin dashed lines). The level of the low $Q$ limits predicted by the model of diffusion inside a sphere are illustrated by shaded areas.[15]

## References

1. J. B. Mietner, F. J. Brieler, Y. J. Lee and M. Fröba, Angewandte Chemie-International Edition **56** (40), 12348-12351 (2017).

2. J. B. Mietner, M. Fröba and R. Valiullin, Journal of Physical Chemistry C **122** (24), 12673-12680 (2018).

3. J. Qvist, H. Schober and B. Halle, Journal of Chemical Physics **134** (14) (2011).

4. S. H. Chen, J. Teixeira and R. Nicklow, Physical Review A **26** (6), 3477-3482 (1982).

5. J. Teixeira, M. C. Bellissentfunel, S. H. Chen and A. J. Dianoux, Physical Review A **31** (3), 1913-1917 (1985).

6. F. Cavatorta, A. Deriu, D. Dicola and H. D. Middendorf, Journal of Physics-Condensed Matter **6**, A113-A117 (1994).

7. D. DiCola, A. Deriu, M. Sampoli and A. Torcini, Journal of Chemical Physics **104** (11), 4223-4232 (1996).

8. M. C. Bellissent-Funel, S. H. Chen and J. M. Zanotti, Physical Review E **51** (5), 4558-4569 (1995).

9. S. Takahara, M. Nakano, S. Kittaka, Y. Kuroda, T. Mori, H. Hamano and T. Yamaguchi, Journal of Physical Chemistry B **103** (28), 5814-5819 (1999).

10. M. Aso, K. Ito, H. Sugino, K. Yoshida, T. Yamada, O. Yamamuro, S. Inagaki and T. Yamaguchi, Pure and Applied Chemistry **85** (1), 289-305 (2013).

11. S. Takahara, N. Sumiyama, S. Kittaka, T. Yamaguchi and M. C. Bellissent-Funel, Journal of Physical Chemistry B **109** (22), 11231-11239 (2005).

12. I. M. Briman, D. Rebiscoul, O. Diat, J. M. Zanotti, P. Jollivet, P. Barboux and S. Gin, Journal of Physical Chemistry C **116** (12), 7021-7028 (2012).

13. D. Laage and J. T. Hynes, Science **311** (5762), 832-835 (2006).

14. J. Qvist, C. Mattea, E. P. Sunde and B. Halle, Journal of Chemical Physics **136** (20) (2012).

15. F. Volino and A. J. Dianoux, Molecular Physics **41** (2), 271-279 (1980).